\documentclass[nofootinbib,reprint,amsmath,amssymb,prd,aps,superscriptaddress,oncolumn]{revtex4-1}
\usepackage{graphicx}
\usepackage{amsmath}
\usepackage{color}
\usepackage{hyperref}
\usepackage{comment}

\begin{document}

\title{Teleparallel dark energy in a nonflat universe}

\author{Rocco D'Agostino}
\email{rocco.dagostino@inaf.it}
\affiliation{INAF -- Osservatorio Astronomico di Roma, Via Frascati 33, 00078 Monte Porzio Catone, Italy}
\affiliation{INFN -- Sezione di Roma 1, Piazzale Aldo Moro 2, 00185 Roma, Italy}

\author{Francesco Bajardi}
\email{f.bajardi@ssmeridionale.it}
\affiliation{Scuola Superiore Meridionale,  Largo San Marcellino 10, 80138 Napoli, Italy}
\affiliation{INFN -- Sezione di Napoli, Via Cintia, 80126 Napoli, Italy}

\begin{abstract}
In this paper, we investigate the cosmological dynamics of teleparallel dark energy in the presence of nonzero spatial geometry. Extending previous analyses of nonminimal scalar-tensor theories in the torsion-based framework, we consider different scalar field potentials and examine the resulting background evolution and linear perturbations. Adopting a dynamical systems approach, we reformulate the field equations and constrain the model parameters via a Markov chain Monte Carlo analysis combining updated datasets from Pantheon+SH0ES supernovae, cosmic chronometers, and growth rate measurements. Our results suggest a mild
preference for an open geometry, although all models remain consistent with a flat universe at the $1\sigma$ level. Notably, Bayesian information criteria indicate that the nonflat teleparallel scenario with a vanishing potential is strongly favored over the standard $\Lambda$CDM model.
Furthermore, all teleparallel scenarios are compatible with local determinations of the Hubble constant and exhibit better agreement with low-redshift structure formation data compared to $\Lambda$CDM. These findings highlight the potential of nonflat teleparallel gravity to address current observational tensions and motivate its further investigation as a viable alternative to standard cosmology.
\end{abstract}

\maketitle

\section{Introduction}
Einstein's theory of general relativity (GR) has been the standard theory of gravity for over a century. Its geometrical interpretation, described in terms of the curvature of spacetime caused by the presence of mass and energy, has successfully explained a wide range of phenomena, from the motion of celestial bodies to the dynamics of large-scale structures in the Universe \cite{Berti:2015itd,LIGOScientific:2016aoc,EventHorizonTelescope:2019dse}. However, despite its success in many domains, GR is not without limitations \cite{Weinberg:1988cp,Padmanabhan:2002ji,Joyce:2014kja,Ishak:2018his,DAgostino:2022fcx}. In particular, it encounters challenges in explaining phenomena at the quantum scale, where the classical nature of GR leads to issues like black hole singularities \cite{Carlip:2001wq,Barack:2018yly}. On the other hand, at the cosmological scales, the standard paradigm grounded on GR -- the $\Lambda$CDM model -- describes the Universe with an unknown amount of cold dark matter (CDM), and a cosmological constant $(\Lambda)$ to account for the enigmatic and theoretically problematic effects of dark energy (DE) \cite{SupernovaCosmologyProject:1998vns,SupernovaSearchTeam:1998fmf,Peebles:2002gy,Copeland:2006wr,DAgostino:2019wko}. 
Furthermore, the $\Lambda$CDM model faces significant discrepancies between early- and late-time measurements of the present rate of the Universe's expansion  \cite{Riess:2019qba,H0LiCOW:2019pvv,DAgostino:2020dhv}, as well as in the amplitude of matter fluctuations \cite{Perivolaropoulos:2021jda,Li:2023azi,DAgostino:2023cgx,DiValentino:2025sru}.

Given these challenges, the search for GR extensions has become a central pursuit in modern physics. These modifications introduce additional degrees of freedom and allow for new gravitational dynamics, which have been particularly useful in explaining the observed accelerated expansion of the Universe \cite{Carroll:2003wy,Starobinsky:2007hu,Sotiriou:2008rp,Nojiri:2010wj,Bajardi:2022tzn,DAgostino:2024sgm,Bajardi:2024efo}. However, very often, simple modifications of the gravitational action 
are not able to fully resolve all the issues that arise at small and large scales.
For this reason, more general theories have been proposed, which incorporate additional fields, such as scalar and vector fields, into the gravitational framework \cite{Clifton:2011jh,Koyama:2015vza,Nojiri:2017ncd,Capozziello:2019cav,Capozziello:2022wgl,DAgostino:2022tdk}. Such extensions aim to relax certain assumptions of GR, providing a more comprehensive approach to understanding gravity at both large and small scales.  Scalar-tensor theories, for example, generalize the concept of gravity by introducing a scalar field alongside the metric, allowing for a richer structure that can potentially explain both cosmological and quantum phenomena more effectively \cite{Brans:1961sx,Horndeski:1974wa,Deffayet:2011gz,Kobayashi:2019hrl,DAgostino:2019hvh}.

Among alternative formulations of gravity, teleparallelism stands out as a compelling and distinctive perspective, in which gravitation is encoded in torsion rather than curvature \cite{Hayashi:1979qx,Maluf:1994ji,Hammond:2002rm,Krssak:2018ywd}. 
In this approach, the gravitational field is described by a flat connection, with torsion serving as the source of gravitational effects. This framework treats gravitational interactions as a force in flat spacetime, analogous to electromagnetism. By imposing a zero-curvature condition and allowing for torsion, teleparallel gravity provides an equivalent but conceptually distinct framework compared to Einstein's theory, known as the teleparallel equivalent to GR (TEGR).
This reformulation not only offers a more direct connection to the dynamics of spacetime but also presents a potential way to address renormalization issues in quantum gravity. Moreover, various extensions such as modified $f(T)$ teleparallel theories have been proposed to explore new insights into DE, cosmic inflation, and the overall dynamics of the Universe \cite{Ferraro:2006jd,Bengochea:2008gz,Linder:2010py,Cai:2015emx,Krssak:2015oua,Abedi:2018lkr,Bahamonde:2021gfp,Bajardi:2024dru}. 
Particularly interesting scenarios are those offered by scalar-tensor teleparallel theories that incorporate both torsion and scalar fields, similar to the GR context, where a scalar field can be added to obtain a dynamical DE sector \cite{Geng:2011aj,Wei:2011yr,Geng:2012vn,DAgostino:2018ngy,Bajardi:2021tul}. This approach potentially offers a viable path to address both the cosmological problems and the small-scale challenges that arise in the context of quantum field theory. 

In this paper, we aim to extend the cosmological analysis of nonminimal scalar-tensor gravity presented in the previous work \cite{DAgostino:2018ngy} by incorporating nonzero spatial curvature into the spacetime metric and investigating the theoretical scenario in light of updated cosmological observations.
The inclusion of spatial curvature plays a key role as it affects not only the background evolution of the Universe but also leaves imprints at the level of perturbations \cite{Yu:2017iju,Capozziello:2018hly,Bahamonde:2022ohm}. At late times, constraining the value of spatial geometry is crucial for understanding the current accelerated expansion of the Universe. At early times, nonzero curvature alters the transfer functions and the power spectra of both tensor and scalar modes generated during inflation \cite{Lewis:1999bs,DAgostino:2023tgm,Califano:2024tns}. Additionally, a deviation from flat geometry can influence the duration of the inflationary phase, which may account for certain features observed in the cosmic microwave background (CMB) power spectrum \cite{Efstathiou:2003hk,Lasenby:2003ur}.
Further motivations to accurately constrain the spatial geometry arise from recent tensions between CMB and baryon acoustic oscillation measurements \cite{Park:2017xbl,DiValentino:2019qzk,Handley:2019tkm}. 
For all these reasons, incorporating the curvature term in modified gravity actions may prove useful in providing a more complete and realistic modeling of cosmic dynamics, as these theories introduce additional degrees of freedom or couplings that interact nontrivially with curvature. This approach could help address the challenges raised by observational data that question the standard assumption of flatness.

The structure of the paper is as follows. In Sec.~\ref{sec:teleparallel}, we outline the main features of teleparallel gravity and present the cosmological solutions to the field equations of a scalar-tensor gravity theory with a nonminimal coupling and nonvanishing spatial curvature. 
In Sec.~\ref{sec:dynamical}, we reformulate the cosmological dynamics of the model in terms of an autonomous system of dimensionless variables, for various forms of the scalar field potential. 
In Sec.~\ref{sec:observational}, we describe the numerical procedure used to compare the theoretical scenarios with cosmic observations at the background and perturbation levels. We then present the constraints on the free parameters of the models and discuss them in light of previous results from complementary probes. Finally, in Sec.~\ref{sec:conclusions}, we conclude by summarizing our findings and suggesting directions for future studies. 

Throughout this work, we set units such that $c=1$ and define $\kappa^2 \equiv 8\pi G$, where $G$ is the Newton constant.

\section{Teleparallel dark energy}
\label{sec:teleparallel}

In the framework of teleparallel gravity, the traditional description of the gravitational interaction based on curvature is replaced in terms of torsion. By relaxing the assumption of symmetric Christoffel connection, it is possible to define the torsion tensor $T^\alpha_{\,\,\,\mu \nu}$ as 
\begin{equation}
\Gamma^\alpha_{\,\,\,[\mu \nu]} := T^\alpha_{\,\,\,\mu \nu}\,.
\label{torsiontens}
\end{equation}
In this way, the most general affine connection that includes both curvature and torsion has the form 
\begin{equation}
\Gamma^\alpha_{\,\,\, \mu \nu} = \hat{\Gamma}^\alpha_{\,\,\, \mu \nu}  + \frac{1}{2}\left(T^{\,\, \alpha}_{\mu \,\,\,\nu} + T_{\mu \nu}^{\,\,\,\alpha} - T^{\alpha}_{\,\, \mu \nu} \right),
\label{generalconnection}
\end{equation} 
where $\hat{\Gamma}^\alpha_{\,\,\, \mu \nu}$ represents the Levi-Civita connection of GR.  

Interestingly, this theory can be recast as a gauge theory for the translation group. This can be achieved by using the tetrad fields $e^a_\mu$ and the spin connection $\omega^a_{\ b \mu}$,  which serve as a link between the flat and the curved space. Tetrad fields form a local orthonormal basis at each point of the spacetime manifold through the relation
\begin{equation}
g_{\mu\nu}=\eta_{ab}e^a_\mu e^b_\nu \,,
\end{equation}
where $\eta_{ab}=\text{diag}(1,-1,-1,-1)$ is the Minkowski metric.
To preserve the gauge invariance and the metric compatibility, the affine connection \eqref{generalconnection} can be written as \cite{Krssak:2015oua}
\begin{equation}
 \Gamma^{\rho}_{\mu\nu} = e_a^{\rho} \left( \partial_\mu e^a_{\nu} + \omega^a_{\ b\mu} \, e^b_{\nu} \right),
\end{equation}
and, therefore, the torsion tensor \eqref{torsiontens} is given by
\begin{equation}
    T^a_{\ \mu\nu}=\partial_\mu e^a_{\nu}-\partial_\nu e^{a}_{\mu}+\omega^a_{\ b\mu}e^b_\nu-\omega^a_{\ b\nu}e^b_{\mu}\,.
\end{equation}
As pointed out in \cite{Krssak:2015oua} and \cite{Hohmann:2018rwf}, the inclusion of the spin connection is fundamental in order to preserve the local Lorentz invariance and avoid frame dependence issues. However, more recent studies \cite{Golovnev:2021lki, Blixt:2022rpl, Golovnev:2023qll} have shown that both the \emph{pure tetrad} formalism (also called the Weitzenb\"ock gauge), in which the spin connection vanishes, and the covariant formalism are dynamically equivalent, yielding the same equations of motion. Nevertheless, the covariant formalism offers a more mathematically refined framework and may facilitate the exploration of non-trivial unconventional matter couplings. In particular, at the level of field equations, one can either work within the pure tetrad formalism or adopt an appropriate tetrad–spin connection pair in the covariant setting. In the Weitzenb\"ock gauge, the affine connection and the torsion tensor  reduce, respectively, to the following expressions:
\begin{align}
     \Gamma^\rho{ }_{\mu \nu}&=e_a^\rho \partial_\mu e_\nu^a\,, \\
   T^\rho_{\,\,\, \mu \nu}&= e_a^\rho \partial_\mu e_\nu^a-e_a^\rho \partial_\nu e_\mu^a\,.
\end{align}
Moreover, the torsion scalar is given by
\begin{equation}
T=S_\rho^{\ \mu\nu}T^{\rho}_{\ \mu\nu}\,,
\label{eq:torsion_scalar}
\end{equation}
where
\begin{align}
    S_{\rho}^{\ \mu\nu}&\equiv \frac{1}{2}\left(K^{\mu\nu}_{\quad \rho}+\delta^\mu_\rho T^{\sigma \nu}_{\quad \sigma}-\delta^\nu_{\rho}T^{\sigma\mu}_{\quad \sigma}\right),\\
    K^{\mu\nu}_{\quad \rho}&\equiv -\frac{1}{2}\left(T^{\mu\nu}_{\quad \rho}-T^{\nu\mu}_{\quad \rho}-T_{\rho}^{\ \mu\nu}\right) ,
\end{align}
which are known as the superpotential and the contortion tensor, respectively. In this framework, the Lagrangian density 
$\mathcal{L}_\text{TEGR}=T$ yields field equations that are dynamically equivalent to those of GR.

In the present study, we extend the TEGR action and consider a scalar-tensor model described by the action 
\begin{equation}
\mathcal{S}=\int d^4x\ e \left[\dfrac{T}{2\kappa^2}+\dfrac{1}{2}\left(\partial_\mu\phi\partial^\mu \phi+\xi T\phi^2\right)-V(\phi)+\mathcal{L}_m\right],
\label{eq:action}
\end{equation}
where a nonminimal coupling between gravity and the scalar field $\phi$ is featured by the dimensionless constant
$\xi$. Here, $V(\phi)$ is the scalar field potential, and $\mathcal{L}_m$ denotes the Lagrangian density of matter fields. This framework is known in the literature as teleparallel dark energy (TDE) \cite{Geng:2011aj}, and represents the torsion analog of nonminimal quintessence in standard GR \cite{Uzan:1999ch}.
The field equations are obtained by varying the action \eqref{eq:action} with respect to the tetrad fields:
\begin{align}
&{\left[e^{-1} \partial_{\mu}\left(e e_a^{\rho} S_{\rho}{ }^{\mu \nu}\right)-e_{a}^{\lambda} T^{\rho}{ }_{\mu \lambda} S_{\rho}{ }^{\nu \mu}-\dfrac{1}{4} e_{a}^{\nu} T\right]\left(\dfrac{2}{\kappa^2}+2 \xi \phi^{2}\right)} \notag\\ 
&+e_{a}^{\mu} \partial^{\nu} \phi \partial_{\mu} \phi+4 \xi e_{a}^{\rho} S_{\rho}{ }^{\mu \nu} \phi \partial_{\mu} \phi-e_{a}^{\nu}\left[\frac{1}{2} \partial_{\mu} \phi \partial^{\mu} \phi-V(\phi)\right]\notag \\ 
&=e_{a}^{\rho} T^{(m)}{ }_{\rho}{ }^{\nu}\,,
\label{eq:FE}
\end{align}
where $T^{(m)}{ }_{\rho}{ }^{\nu}$ is the stress-energy tensor of matter.

\subsection{Background cosmology}
We conduct our study within the cosmological background described by the homogeneous and isotropic Friedmann-Lema\^itre-Robertson-Walker (FLRW) metric:
\begin{equation}
    ds^2=dt^2-a^2(t)\left[\frac{dr^2}{1-kr^2}+r^2\left(d\theta^2+\sin^2\theta\, d\varphi^2\right)\right],
    \label{eq:metric}
\end{equation}
where $a(t)$ is the normalized cosmic scale factor, and $k$ is the spatial curvature parameter\footnote{In our formalism, $a_0=1$ is the present value of the scale factor, while $k=\left\{-1,0,1\right\}$ defines the geometry of an open, flat and closed universe, respectively.}. 
It is worth noting that, unlike in GR where the connection is of Levi-Civita type and thus uniquely determined by the metric, in teleparallel gravity, a given line element can be realized in different ways \cite{Hohmann:2020zre}. In fact, as discussed in the previous section, spacetime~\eqref{eq:metric} depends on the specific choice of both the tetrad and the spin connection. Since our focus here is on analyzing the field equations, a practically and commonly adopted approach is to work within pure tetrad formalism, where the tetrad takes the diagonal form 
\begin{align}
    &e^a_{\ \mu}= \text{diag} \left( 1,\, \frac{a(t)}{\sqrt{1 - k r^2}},\, a(t) r,\, a(t) r \sin\theta \right).
    \label{eq:tetrads}
\end{align}
By pursuing this approach, the resulting field equations are dynamically equivalent to those obtained when an appropriate spin connection is included, as demonstrated in \cite{Krssak:2015oua} for modified $f(T)$ gravity. Specifically, it was shown that there exists a set of spin connection whose nonvanishing components are given by
\begin{eqnarray}
    &&\omega^1_{\ 2\theta}=-\sqrt{1-kr^2}\,, \quad \omega^1_{\ 3\varphi}=-\sqrt{1-kr^2}\sin\theta\,, \nonumber
    \\
    &&\omega^2_{\ 3\varphi}=-\cos\theta\,,
\end{eqnarray}
which provide the same field equations as in the pure tetrad case. In our setup, using the tetrad fields \eqref{eq:tetrads}, Eq.~\eqref{eq:torsion_scalar} becomes
\begin{equation}
T = - 6\left(H^2 -\frac{k}{a^2}\right),
\label{Torsion}
\end{equation}
where $H \equiv \dot{a}/a$ is the Hubble expansion rate. Then, Eqs.~\eqref{eq:FE} yield the modified Friedmann equations
\begin{align}
   &H^2 + \frac{k}{a^2} = \frac{\kappa^2}{3} \left(\rho_m +\rho_{\phi} \right), \label{eq:Friedmann1} \\
   &\dot{H}=-\dfrac{\kappa^2}{2}\left(\rho_m+\rho_\phi+p_\phi\right), \label{eq:Friedmann2}
\end{align}
where  $\rho_m$ is the matter density governed by the continuity equation 
\begin{equation}
    \dot{\rho}_m+3H\rho_m=0\,. 
    \label{eq:continuity}
\end{equation}
In this framework, the DE contribution is expressed in terms of an effective fluid, whose density and pressure are those associated with $\phi$ and are given by, respectively, 
\begin{align}
&\rho_\phi=\dfrac{1}{2}\dot{\phi}^2+V(\phi)-3H^2\xi\phi^2-\frac{3k}{a^2}\xi\phi^2\,, \label{eq:rho_phi}\\
&p_\phi=\dfrac{1}{2}\dot{\phi}^2-V(\phi) +4H\xi\phi\dot{\phi}+\left(3H^2+2\dot{H}\right)\xi\phi^2+\frac{2 k}{a^2}\xi  \phi ^2\,.
\label{eq:p_phi}
\end{align}
As it can be seen from the last two terms on the right-hand side of  Eqs.~\eqref{eq:rho_phi} and \eqref{eq:p_phi}, the nonzero curvature also contributes to the effective DE fluid through the coupling constant $\xi$. Consistently, for $k\to 0$, the cosmic dynamics reduces to that previously investigated in \cite{DAgostino:2018ngy}.

We can then define the effective DE equation of state (EoS) parameter, $w_\text{DE}\equiv p_\phi/\rho_\phi$, which satisfies the continuity equation
\begin{equation}
    \dot{\rho_\phi}+3H(1+w_\phi)\rho_\phi=0\,.
\end{equation}
Additionally, the equation of motion for the scalar field is obtained by varying Eq.~\eqref{eq:action} with respect to $\phi$:
\begin{equation}
    \ddot{\phi}+3H\dot{\phi}+V_\phi=\xi T \phi\,,
    \label{eq:Klein-Gordon}
\end{equation}
where $V_\phi\equiv dV/d\phi$. It is worth noting that standard quintessence is recovered in the limit $\xi\to 0$. 
\subsection{Linear matter perturbations}

The study of linear perturbations plays a crucial role in distinguishing between GR and modified gravity theories. While background cosmological evolution provides valuable insights into the Universe's expansion history, it often fails to uniquely identify deviations from GR, as several models can produce similar dynamics. 
In fact, any modifications of gravity affect not only the background dynamics, but also the way matter clusters under gravitational attraction.  
Therefore, analyzing the evolution of density fluctuations could help to break degeneracies that arise when considering only background observables.

The growth of structure in the Universe can be examined through the evolution of the density contrast $\delta\equiv \delta\rho_m/\rho_m$, which depends on the underlying gravitational theory \cite{Peebles:1980yev,1993sfu..book.....P}. In particular, on subhorizon scales and within the linear regime, we have
\begin{equation}
    \ddot{\delta}+2H\dot{\delta}_m-4\pi G_\text{eff}\rho_m\delta=0\,,
    \label{eq:delta_m}
\end{equation}
where $G_\text{eff}$ is the effective gravitational coupling, which measures the deviations from the Newton constant. 
Then, the evolution of matter density fluctuations can be conveniently rewritten using Eq.~\eqref{eq:delta_m} in terms of the derivatives with respect to the scale factor:
\begin{equation}
  \frac{d^2\delta}{da^2}+ \left(\frac{3}{a}+\frac{d\ln H}{da}\right)\frac{d\delta}{da}=\frac{4\pi G_\text{eff}}{a^3 H^2} \delta\,.
  \label{eq:matter_perturbations_1}
\end{equation}
Although the evolution equation for matter fluctuations is formally equivalent to that of a flat universe, the effects of the nonzero curvature are encoded in $H(a)$ through Eq.~\eqref{eq:Friedmann1} and influence the initial conditions for the growth of perturbations. Consequently, the presence of spatial curvature alters the transition phase between the different cosmic eras.

Deviations from the Newton constant can be quantified through  \cite{Geng:2012vn,Abedi:2018lkr}
\begin{equation}
    G_\text{eff}=\frac{G}{1+\xi\kappa^2\phi^2}\left[1-\frac{\kappa^2\dot\phi^2}{2H^2(1+\xi\kappa^2\phi^2)}\right].
    \label{eq:Geff}
\end{equation}
For $\xi=0$, the kinetic energy of the scalar field becomes subdominant compared to the DE at the present time, and so the Newton constant is recovered. 
It is worth noting that Eq.~\eqref{eq:Geff} was originally obtained in the context of spatially flat models, based on the linear scalar perturbation equations under the quasi-static approximation. However, the underlying assumptions, namely the dominance of subhorizon modes and slowly evolving background fields, remain valid for spatially curved FLRW backgrounds as long as the curvature scale is much larger than the perturbation scale. In this regime, spatial curvature contributes negligibly to the Poisson equation on subhorizon scales, allowing the same expression for $G_\text{eff}$ to hold to leading order.

\section{Autonomous dynamical system}
\label{sec:dynamical}

To solve the cosmological dynamics of the model, we apply the method of dynamical systems (see \cite{Bahamonde:2017ize} for a review), where one can recast the cosmological equations into an autonomous system of dimensionless variables. For this purpose, we define\footnote{A direct comparison with the results presented in \cite{DAgostino:2018ngy} can be done through the mapping $\{x_1,x_2,x_3,x_4,x_5\}\rightarrow \{x,y,v,u,1\}$.}
\begin{subequations}
\begin{align}
    &x_1:=\frac{\kappa \dot{\phi}}{6H^2}\,, \quad x_2:=\frac{\kappa\sqrt{V}}{\sqrt{3}H}\,, \quad x_3:=\frac{\kappa \sqrt{\rho_m}}{\sqrt{3}H}\,, \\
    &x_4:=\kappa \phi\,, \quad x_5:=\sqrt{1+\frac{k}{a^2H^2}}\,,
\end{align}
 \label{eq:variables}
\end{subequations}
such that, Eq.~\eqref{eq:Friedmann1} reads as
\begin{equation}
    x_1^{\,2}+x_2^{\,2}+x_3^{\,2}=x_5^{\,2}(1+\xi x_4^{\,2})\,.
    \label{eq:sum}
\end{equation}
Moreover, using Eq.~\eqref{eq:Friedmann2}, we define
\begin{equation}
    s:=-\frac{\dot{H}}{H^2}=1-x_5^{\,2}+\frac{6 x_1^{\,2}+4 \sqrt{6}\, \xi  x_1 x_4+3 x_3^{\,2}}{2 \left(1+\xi x_4^{\,2}\right)}\,.
    \label{eq:s}
\end{equation}
Therefore, the dynamics of the model can be analyzed in terms of derivatives of the variables  \eqref{eq:variables} with respect to the $e$-folding number, $N\equiv \ln a$. In particular, using Eqs.~\eqref{eq:continuity} and \eqref{eq:Klein-Gordon}, we obtain
\begin{subequations}
\begin{align}
x_1'&= (s-3)x_1+\sqrt{6}\, \xi x_4 \left(x_5^{\,2}-2\right) -\frac{\kappa\,  V_\phi}{\sqrt{6} H^2} \,, \label{eq:x1'}\\
x_2'&=s\,x_2+ \frac{x_1 V_\phi}{\sqrt{2V} H}\,, \label{eq:x2'}\\
x_3'&=  -\frac{x_3}{2} \,, \\
x_4'&=\sqrt{6}\, x_1\,, \label{eq:x4'} \\
x_5'&=\frac{(s-1) \left(x_5^{\,2}-1\right)}{x_5}\,.  \label{eq:x5'}
\end{align}
\label{eq:system}
\end{subequations}
It is worth noting that one of the above equations may be eliminated in view of the constraint given by Eq.~\eqref{eq:sum}.

To provide a more physical interpretation of the dimensionless variables, we introduce the normalized energy density parameters of the cosmic species. For matter and curvature we have, respectively,
\begin{equation}
    \Omega_m:=\frac{\kappa^2 \rho_m}{3H^2}=x_3^{\,2}\,, \quad \Omega_k:=\frac{-k}{a^2H^2}=1-x_5^{\,2}\,,
\end{equation}
while the DE effects are quantified by the barotropic fluid parameter
\begin{equation}
    w_\text{DE}=\frac{3 x_1^{\,2} - 3 x_2^{\,2} + 4 \sqrt{6}\,\xi x_1 x_4 +\xi x_4^{\,2}(1 - 2 s + 2 x_5^{\,2})  }{3 \left(x_1^{\,2}+x_2^{\,2}-\xi x_4^{\,2} x_5^{\,2}\right)}\,.
    \label{eq:barotropic}
\end{equation}
Finally, using Eq.~\eqref{eq:Geff}, we can express the deviations from Newton's gravitational constant as
\begin{equation}
   \frac{G_\text{eff}}{G}= \frac{1-3x_1^{\,2}+\xi x_4^{\,2}}{(1+\xi x_4^{\,2})^2}\,.
   \label{eq:Geff_2}
\end{equation}

In order to close the system of Eqs.~\eqref{eq:system}, we need to specify the scalar field potential. To this end, we analyze three different scenarios, as outlined in the following subsections. 

\subsection{Vanishing potential}

We first consider the zero-potential case:
\begin{align}
    V(\phi)=0\,.
\end{align}
Here, the dynamics is purely governed by the scalar field and its coupling to the scalar torsion.
Then, $x_2=0$, and the autonomous system simplifies as
\begin{subequations}
\begin{align}
    x_1'&=(s-3)x_1+\sqrt{6}\,\xi x_4\,, \\
    x_4'&=\sqrt{6}\, x_1\,,\\
    x_5'&=\frac{(s-1) \left(x_5^{\,2}-1\right)}{x_5}\,, 
\end{align}
\end{subequations}
while, from Eq.~\eqref{eq:sum}, we find
\begin{equation}
    x_3=\sqrt{x_5^{\,2} \left(1+\xi x_4^{\,2}\right)-x_1^{\,2}}\,.
\end{equation}

\subsection{Exponential potential}

To explore a more general and realistic scenario, we extend the analysis by considering an exponential form for the scalar field potential: 
\begin{equation}
    V(\phi)=V_0\,e^{-\kappa \phi}\,,
\end{equation}
where $V_0$ is a normalization constant.
In this case, the autonomous system takes the form
\begin{subequations}
\begin{align}
    x_1'&=(s-3) x_1-\sqrt{\frac{3}{2}}\, x_2^{\,2}+\sqrt{6}\,\xi x_4(x_5^{\,2}-2)\,, \\
    x_2'&=s\, x_2+\sqrt{\frac{3}{2}}\, x_1 x_2\,,\\
    x_4'&=\sqrt{6}\, x_1\,,  \\
    x_5'&=\frac{(s-1) \left(x_5^{\,2}-1\right)}{x_5}\,,
\end{align}
\end{subequations}
subject to the constraint given by Eq.~\eqref{eq:sum}.

\subsection{Power-law potential}

Furthermore, we introduce an additional degree of freedom in the theoretical setup, allowing for a broader range of dynamical behavior. In particular, we focus on the power-law potential 
\begin{equation}
    V(\phi)=V_0(\kappa\phi)^n\,,
\end{equation}
where $n\geq0$. Therefore, the autonomous system reads 
\begin{subequations}
\begin{align}
    x_1'&=(s-3)x_1-\sqrt{\frac{3}{2}}\,\frac{n\, x_2^{\,2}}{x_4}+\sqrt{6}\,\xi x_4(x_5^{\,2}-2)\,, \\
    x_2'&=s\,x_2+\sqrt{\frac{3}{2}}\,\frac{n\,x_1 x_2}{x_4}\,, \\
    x_4'&=\sqrt{6}\, x_1\,, \\
    x_5'&=\frac{(s-1) \left(x_5^{\,2}-1\right)}{x_5}\,,
\end{align}
\end{subequations}
while Eq.~\eqref{eq:sum} still holds.

\begin{table*}
\begin{center}
\setlength{\tabcolsep}{0.5em}
\renewcommand{\arraystretch}{2}
\begin{tabular}{c c c c c c c c} 
\hline
\hline
Model & $h$ & $\Omega_{m,0}$ & $\Omega_{k,0}$ & $\xi$  & $n$ & $\sigma_{8}$  \\
\hline 
$V=0$ & $0.717^{+0.009(0.017)}_{-0.009(0.017)}$ & $0.333^{+0.027(0.052)}_{-0.027(0.052)}$ & $0.003^{+0.030(0.050)}_{-0.029(0.051)}   $ & $-0.342^{+0.006(0.015)}_{-0.008(0.014)}$ & -- & $0.760^{+0.118(0.219)}_{-0.112(0.224)}      $ \\
$V=V_0e^{-\kappa\phi}$ & $0.716^{+0.009(0.017)}_{-0.009(0.017)}$ & $0.326^{+0.021(0.045)}_{-0.022(0.043)}   $ & $0.002^{+0.008(0.016)}_{-0.008(0.017)} $ & $-0.345^{+0.005(0.010)}_{-0.005(0.010)}$ & -- & $0.753^{+0.046(0.093)}_{-0.049(0.084)}   $\\
$V=V_0(\kappa\phi)^n$ & $0.718^{+0.009(0.017)}_{-0.009(0.017)}   $ & $0.298^{+0.026(0.049)}_{-0.024(0.052)}   $ & $0.009^{+0.012(0.028)}_{-0.015(0.025)}   $ &  $-0.352^{+0.015(0.025)}_{-0.011(0.030)}  $ & $0.95^{+0.14(0.23)}_{-0.11(0.25)}  $ & $0.776^{+0.046(0.107)}_{-0.054(0.098)}   $ \\
\hline
\hline
\end{tabular}
\caption{68\% (95\%) limits on the free parameters of the nonflat TDE models, obtained from the MCMC analysis on the Pantheon+SH0ES+CC+GRF data.}
 \label{tab:results}
\end{center}
\end{table*}

\begin{figure*}
    \centering
    \includegraphics[width=3.3in]{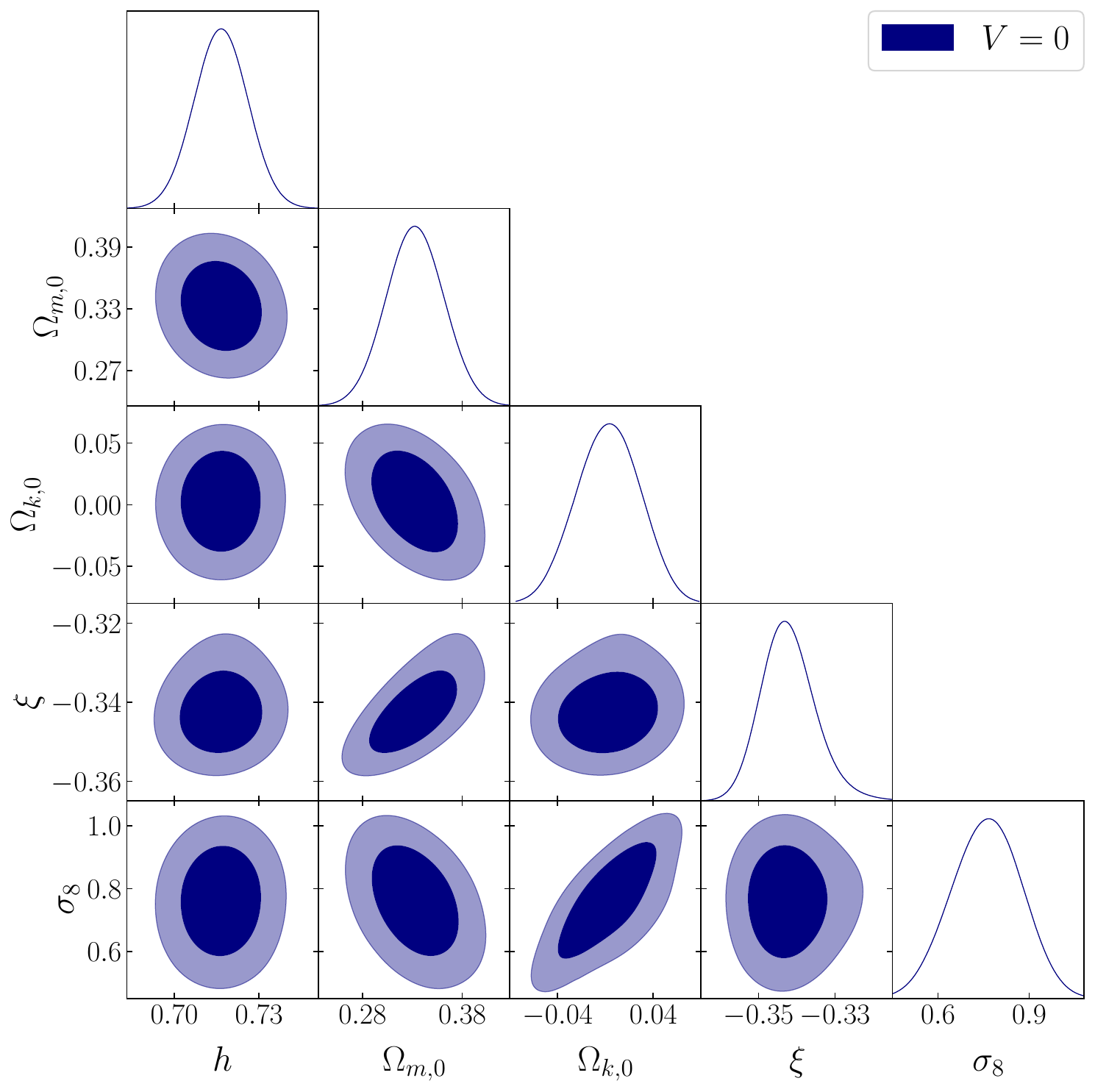}\qquad
    \includegraphics[width=3.3in]{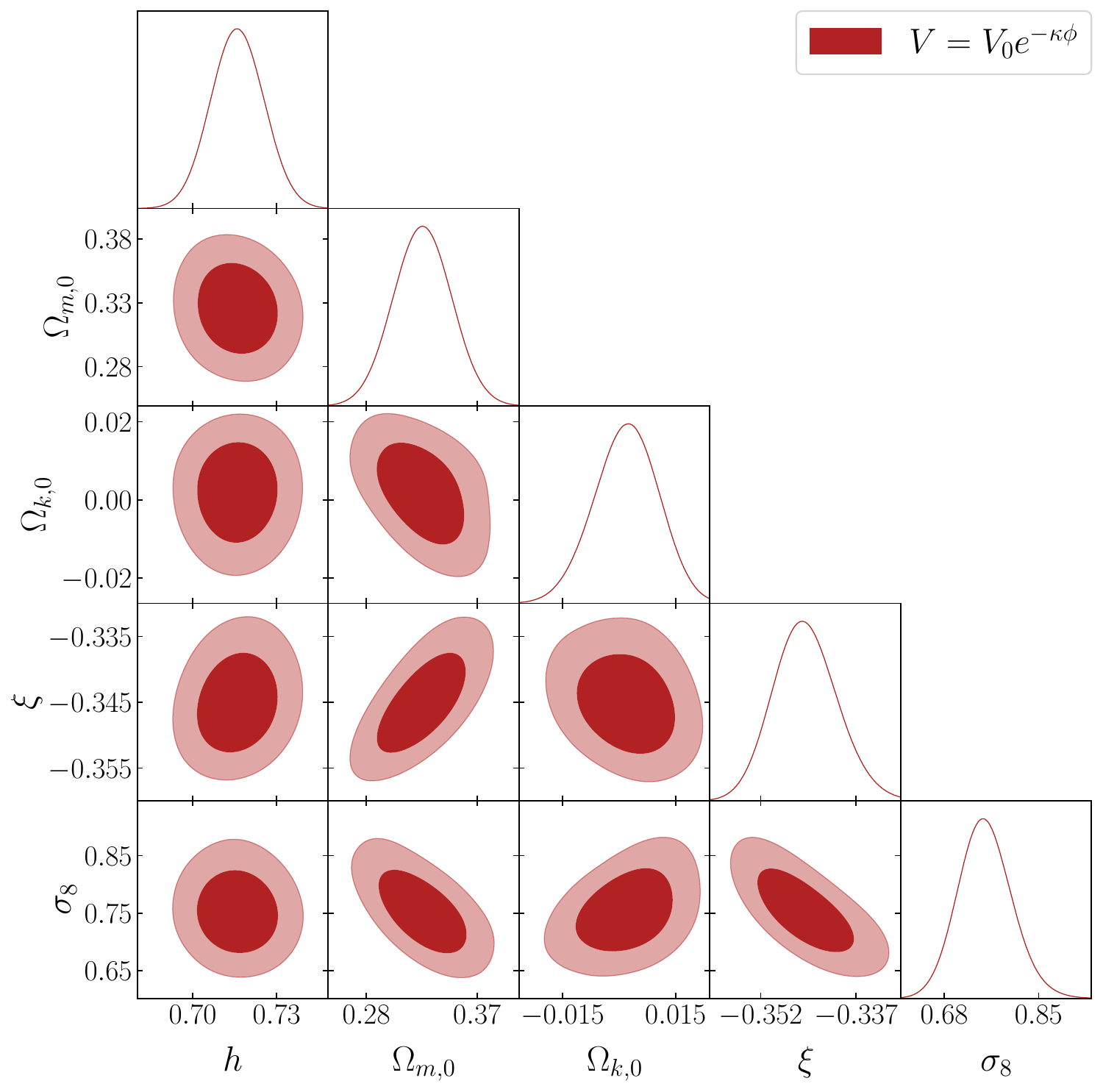}\\
    \vspace{0.2cm}
    \includegraphics[width=3.3in]{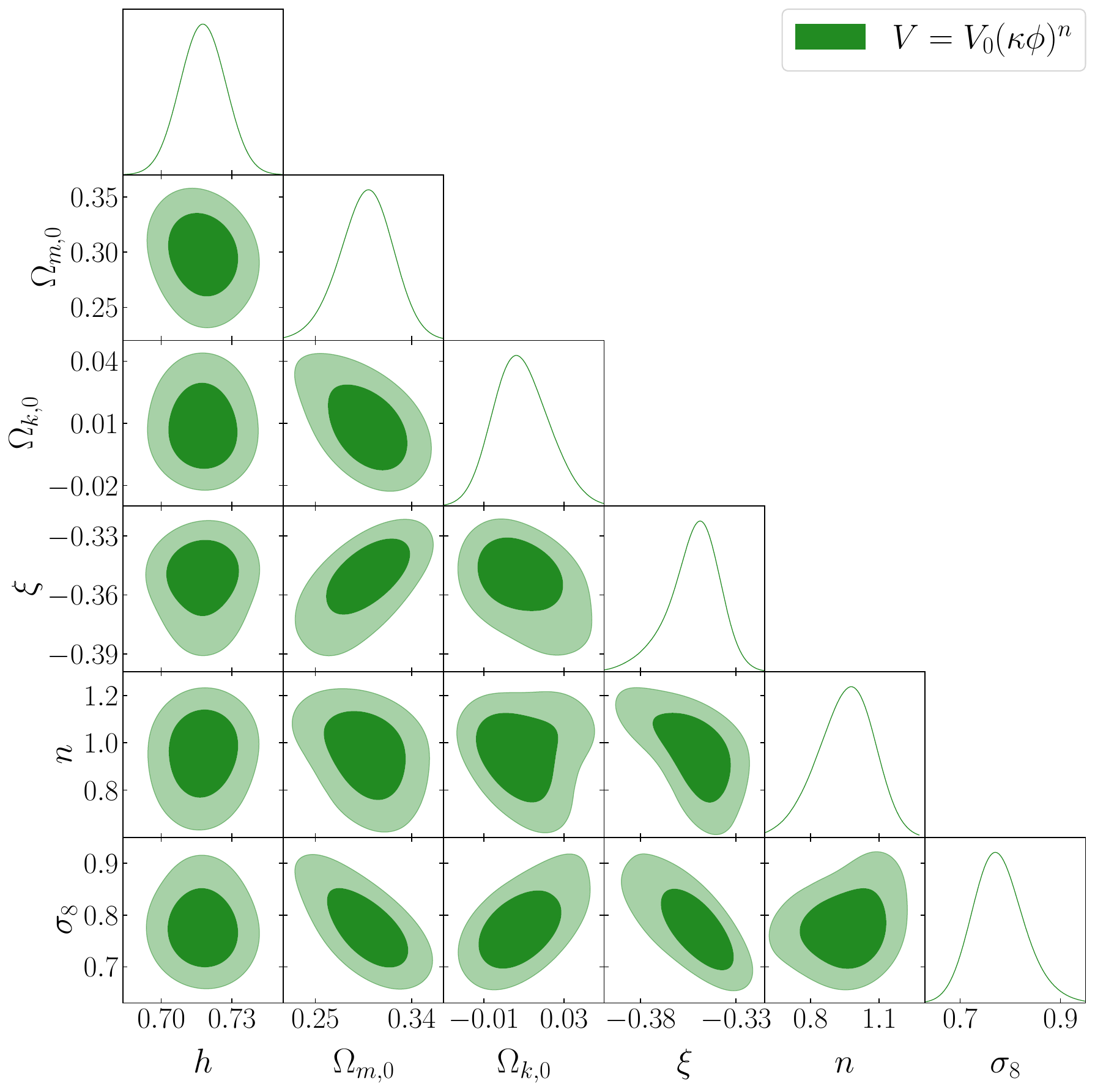}
    \caption{$68\%$ and $95\%$ regions of the nonflat TDE models, along with posterior distributions, obtained from the MCMC analysis on the Pantheon+SH0ES+CC+GRF data.}
    \label{fig:contours}
\end{figure*}

\section{Observational tests}
\label{sec:observational}

The theoretical scenarios presented above shall now be tested against cosmological observations. 
We note that, once the evolution of the dimensionless variables is determined,
the DE effects can be inferred from Eq.~\eqref{eq:barotropic}. Thereby, the background dynamics is given as
\begin{equation}
H(a)=H_0\left[\frac{\Omega_{m,0}}{a^3}+\frac{\Omega_{k,0}}{a^2}+\Omega_\text{DE,0}\,g(a)\right]^{1/2},
\end{equation}
where $H_0$ is the Hubble constant. Here, $\Omega_\text{DE,0}=1-\Omega_{m,0}-\Omega_{k,0}$ is the present DE density, while we have
\begin{equation}
    g(a)\equiv \exp\left\{-3\int_a^1\left[\frac{1+w_\text{DE}(\tilde{a})}{\tilde{a}}\right]d\tilde{a}\right\}.
\end{equation}
We probe the background evolution through the most recent sample of type Ia supernova (SN) data, the Pantheon+ set \cite{Scolnic:2021amr,Brout:2022vxf}, consisting of 1701 light curves of 1550 SNe in the redshift\footnote{The redshift is conventionally defined as $z\equiv a^{-1}-1$.} interval $z\in[0.001,\, 2.26]$,  in combination with the SH0ES Cepheid host distances \cite{Riess:2021jrx}\footnote{The complete Pantheon+SH0ES dataset and the corresponding covariance matrix are available at \url{https://github.com/PantheonPlusSH0ES/DataRelease}.}. In particular, we adopt the analytical treatment presented in \cite{SNLS:2011lii} to marginalize over the SN absolute magnitude and remove it from the fitting analysis. Additionally, we combine the SN data with the cosmic chronometer (CC) measurements acquired through the differential age method \cite{Jimenez:2001gg}. In particular, we use the independent sample of 31 observational Hubble data collected in \cite{Capozziello:2017buj} over the redshift range $z\in[0.07,1.965]$.

Additionally, we probe the evolution of density perturbations using Eq.~\eqref{eq:matter_perturbations_1} in the form
\begin{equation}
    \frac{d^2\delta(a)}{da^2}+\left[\frac{3}{a}+\frac{d\ln E(a)}{da}\right]\frac{d\delta(a)}{da}-\frac{3\Omega_{m,0}G_\text{eff}(a)}{2a^5E(a)^2G}\delta(a)=0\,,
    \label{eq:matter_perturbations_2}
\end{equation}
where $E(a)\equiv H(a)/H_0$ is the normalized Hubble rate, while the ratio $G_\text{eff}/G$ is computed from Eq.~\eqref{eq:Geff_2}. 
In this case, we use the compilation of 18 growth rate factor (GRF) data presented in \cite{Nesseris:2017vor} over the redshift range $z\in [0.02,1.40]$. These are independent measurements from various weak lensing and redshift space distortion surveys of the observable $f\sigma_8(a)\equiv f(a)\sigma_8(a)$, where 
\begin{equation}
    f(a)\equiv \frac{d\ln \delta(a)}{d\ln a}\,,\quad \sigma_8(a)\equiv\sigma_8\frac{\delta(a)}{\delta(a=1)}\,,
\end{equation}
with $f(a)$ denoting the growth rate of matter density perturbations \cite{Wang:1998gt,Linder:2005in} which depends on the specific theory of gravity. Moreover, the quantity $\sigma_8(a)$ measures the amplitude of linear density fluctuations within spheres of radius $8h^{-1}$Mpc, where $\sigma_8$ refers to its value at the present time, and $h\equiv H_0$/(km/s/Mpc) is the dimensionless Hubble constant. It should be noted that the GRF data are model-dependent measurements and, thus, should be corrected for the fiducial cosmology. To do this, we follow the procedure described in \cite{Nesseris:2017vor} and remove the bias induced by the assumed $\Lambda$CDM model.

Therefore, we apply the Metropolis-Hastings algorithm \cite{Hastings70} to perform a Markov chain Monte Carlo (MCMC) analysis of the joint likelihood function
\begin{equation}
    L_\text{joint}=L_\text{Pantheon+SH0ES}\times L_\text{CC}\times L_\text{GRF}\,,
\end{equation}
which includes a total of 1750 data points.
Following the strategy of \cite{DAgostino:2018ngy}, we numerically integrate the system \eqref{eq:system} by setting the initial conditions at $a_\text{in}=10^{-2}$, such that $x_{2,\text{in}}=10^{-6}$, $x_{3,\text{in}}=\sqrt{\Omega_{m,\text{in}}}$\,, $x_{4,\text{in}}=10^{-6}$ and $x_{5,\text{in}}=\sqrt{1-\Omega_{k,\text{in}}}$, with $\Omega_{m,\text{in}}=0.999$ and $\Omega_{k,\text{in}}=10^{-6}$. The boundary condition on $x_1$ can be obtained by inverting Eq.~\eqref{eq:sum} at $a=a_\text{in}$.
Furthermore, we assume flat priors on the free parameters: $h\in[0.6,0.8]$, $\Omega_{m,0}\in[0,1]$, $\Omega_{k,0}\in [-0.1,0.1]$, $\xi\in[-1,0]$\, $n\in[0,2]$ and $\sigma_8\in [0.5,1.2]$\,.
It is worth noticing that Eq.~\eqref{eq:matter_perturbations_2} does not depend on $H_0$, and thus the numerical analysis of the GRF data is insensitive to $h$. 

As a benchmark, we also investigate the predictions of the standard FLRW cosmology. The corresponding results are reported in the Appendix.

\subsection{Numerical results}
\label{sec:numerical}

The results of our MCMC analysis are summarized in Table~\ref{tab:results}, where we present the mean values of the free parameters for the different nonflat TDE scenarios, along with the corresponding limits up to the $2\sigma$ confidence level. 
Furthermore, in Fig.~\ref{fig:contours}, we display the marginalized $1\sigma$ and $2\sigma$ contour plots along with the posterior distributions.
We note that our results show small mean values of $\Omega_{k,0}>0$ across all models, indicating a slight preference for an open universe. However, all TDE scenarios remain consistent with a flat universe within the $1\sigma$ confidence level. 
The largest mean curvature is exhibited by the power-law potential model, which deviates by $\sim 1\%$ from a perfectly flat universe.

The coupling between the scalar field and the scalar torsion remains consistently negative across all teleparallel models, with mean values in the range $-0.35\lesssim\xi\lesssim -0.34$. This indicates a stable and significant interaction, marking a clear deviation from minimally coupled teleparallel quintessence. We also observe that the power-law potential model has a slightly broader $1\sigma$ uncertainty on $\xi$ $(\sim 4\%)$ compared to the the null potential $(\sim 2\%)$ and exponential models $(\sim 1\%)$, reflecting greater flexibility in parameter space due to the presence of an additional degree of freedom. 
It is worth noticing that, in the TDE scenario with a power-law potential, the smaller matter density is associated with a greater absolute magnitude of $\xi$, meaning the additional degrees of freedom shift the matter density to balance the modified gravitational dynamics introduced by the torsion-scalar coupling.
Notably, our results for the power index are fully consistent with $n=1$ at the $1\sigma$ level, suggesting that a simple linear potential may be a viable explanation within the teleparallel framework.

In Fig.~\ref{fig:wDE}, we use the mean results of the MCMC analysis to reconstruct the late-time behavior of the effective EoS parameter of DE. We observe that all scenarios can cross the phantom line divide, specifically at $z\simeq\{0.10,0.14,0.32\}$ for the null potential, exponential potential and power-law potential models, respectively. At the present time, we have $w_\text{DE}\simeq \{-1.12,-1.17,-1.27\}$ for the same models, respectively.

\subsection{Comparison with Planck-$\Lambda$CDM cosmology}
\label{sec:Planck}

It is now interesting to compare our findings with the final results of the Planck Collaboration based on the CMB measurements within the standard $\Lambda$CDM cosmology \cite{Planck:2018vyg}. In what follows, when quoting the Planck results, we specifically refer to the TT,TE,EE+lowE+lensing measurements at the 68\% confidence level.

Regarding the density parameters, Planck predicts $\Omega_{m,0}=0.315\pm 0.007$. This result is in agreement within $1\sigma$ with the outcomes of all TDE models.
Moreover, when allowing for nonzero curvature, the Planck results yield $\Omega_{k,0}=-0.0106\pm 0.0065$, which is consistent with a flat universe only within $2\sigma$. In this respect, the mean Planck results slightly favor a closed universe in contrast with the predictions of the nonflat TDE models (see Sec.~\ref{sec:numerical}).
In particular, the Planck prediction differs by $\sim 1.3\sigma$ from the TDE scenarios with exponential and power-law potentials. Conversely, our results for the TDE model with vanishing potential, as well as for the standard cosmology (see Table~\ref{tab:results_standard} and Fig.~\ref{fig:contours_standard}), are consistent with the Planck predictions at $1\sigma$.

\begin{figure*}
    \centering
    \begin{minipage}{0.48\textwidth}
    \includegraphics[width=3.2in]{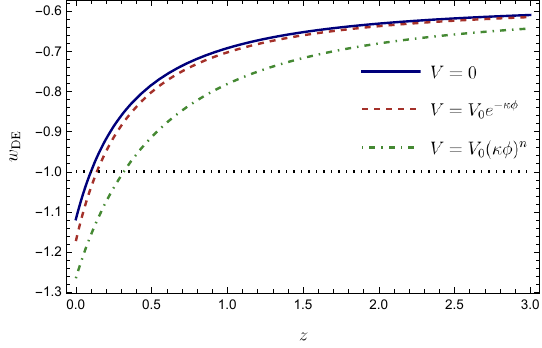}
    \caption{Redshift evolution of the effective DE EoS parameter based on the mean values of the MCMC results for the nonflat TDE scenarios of the present study. The dotted line indicates the phantom divide.}
    \label{fig:wDE}
    \end{minipage}
    \hfill 
    \begin{minipage}{0.48\textwidth}
    \includegraphics[width=3.2in]{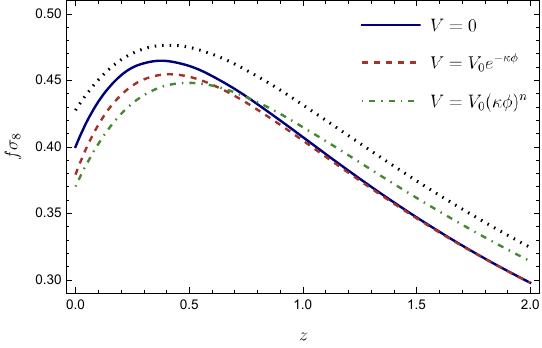}
    \caption{Growth rate of fluctuations as a result of the mean values of the MCMC analysis of the nonflat TDE scenarios of the present study. The dotted curve refers to the Planck-$\Lambda$CDM cosmology (see discussion in Sec.~\ref{sec:Planck}).}
    \label{fig:fsigma8}
    \end{minipage}
\end{figure*}

Within the framework of GR, the parameter values derived from the Planck-$\Lambda$CDM model suggest a systematically stronger growth of cosmological perturbations than what is observed through dynamical probes.
This tension is commonly quantified through the parameter $S_8\equiv \sigma_8\sqrt{\Omega_{m,0}/0.3}$. In particular, CMB measurements predict $S_8=0.832\pm 0.013$, which is in $\sim 2-3\sigma$ tension with cluster counts, redshift space distortion and weak lensing observations (see \cite{Perivolaropoulos:2021jda} for a review). 
To compare these results with our findings, we use the values in Table~\ref{tab:results} and obtain $S_8=0.801 \pm 0.150$, $ S_8=0.785 \pm 0.072$, and $S_8=0.773 \pm 0.075$ for the TDE models with vanishing, exponential and power-law potentials, respectively. As shown in Fig.~\ref{fig:fsigma8}, these values are all lower than the Planck result and more compatible with local probes, although uncertainties -- especially for the zero potential case -- are relatively larger.
Therefore, our results can be interpreted as hints that teleparallel modifications of gravity could help explain current cosmological tensions, suggesting that the standard $\Lambda$CDM model might need to be extended.

Finally, we want to compare our results for the Hubble constant with the Planck prediction, $h=0.674\pm 0.005$. This result shows significant deviations $(>4\sigma)$ from those obtained in all nonflat TDE scenarios under examination. Vice versa, our results are in agreement at the $1\sigma$ level with the model-independent estimate based on local Cepheids calibrations by the SH0ES team \cite{Riess:2021jrx}, $h=0.73 \pm 0.01$. 
This discrepancy with the Planck value, together with the natural alignment of TDE models with the SH0ES measurements, may point toward new physics beyond the standard $\Lambda$CDM paradigm. 

\begin{table}
\begin{center}
\setlength{\tabcolsep}{0.8em}
\renewcommand{\arraystretch}{2}
\begin{tabular}{c c c c} 
\hline
\hline
Model & $\Delta$AIC  & $\Delta$DIC  \\
\hline 
$V=0$ & $-10.4$  & $-14.1$\\
$V=V_0e^{-\kappa\phi}$ & 2.11 & 0.30\\
$V=V_0(\kappa\phi)^n$ & 2.84  & $-0.70$\\
\hline
\hline
\end{tabular}
\caption{Bayesian selection criteria for the nonflat TDE scenarios relative to the $\Lambda$CDM model.}
 \label{tab:Bayes}
\end{center}
\end{table}

\subsection{Bayesian evidence}

To evaluate the predictivity power of different models, it is useful to employ statistical descriptors that measure the Bayesian complexity across different prior volumes. 
Valuable tools for comparing models are offered, for example, by the Akaike information criterion (AIC) \cite{Akaike:1974vps}  and deviance information criterion (DIC) \cite{Spiegelhalter:2002yvw}. 
These indicators are widely used in astrophysical contexts to identify models that not only fit the data well but also avoid unnecessary complexity. Unlike a simpler $\chi^2$ analysis\footnote{Note the relation $\chi^2=-2\ln L$.}, these methods penalize extra degrees of freedom, helping to quantify the evidence in favor of simpler, more predictive models. Specifically, we have
\begin{align}
\text{AIC}&\equiv-2\ln L_\text{max}+2p\,, \\
\text{DIC}&\equiv 2\langle -2\ln L\rangle+2\ln\langle L\rangle\,,
\end{align}
where $p$ is the number of independent parameters within a model, $L_\text{max}$ is the maximum likelihood value, and $\langle\cdot\rangle$ indicates the mean over the posterior distribution. 
It is worth noting that the AIC assumes that all parameters are well-constrained by the data and penalizes extra parameters accordingly. However, this might not always be the case, and the DIC parameter appears sometimes more suitable since it takes into account the actual effective number of parameters as measured by the Bayesian complexity \cite{Trotta:2008qt,Liddle:2007fy}.

In order to interpret the Bayesian evidence, it is conventional to compute the differences $(\Delta$) in the values of information criteria between two models, where the best model is the one characterized by the lowest $\Delta$. 
The significance is then assessed based on Jeffrey's scale as follows \cite{Kass:1995loi}: 
\begin{itemize}
\item $|\Delta|\in(0,2)$: weak evidence; 
\item $|\Delta|\in (2,6)$: substantial evidence; 
\item $|\Delta|\in(6,10)$: strong evidence; 
\item $|\Delta|>10$: decisive evidence.
\end{itemize}

In Table~\ref{tab:Bayes}, we present the $\Delta$AIC and $\Delta$DIC values relative to the standard $\Lambda$CDM model, which is characterized by the lowest number of free parameters (see Appendix). 
Remarkably, we can observe that the TDE model with vanishing potential is very strongly preferred against the standard scenario, while the model with an exponential potential shows only weak evidence in its favor. Moreover, we note that the discrepancy between the $\Delta$AIC and $\Delta$DIC values for the TDE scenario with a power-law potential arises from their different interpretations of the model complexity. In this case, the three additional parameters compared to the benchmark model are heavily penalized by the AIC, which favors the simpler scenario. In contrast, the DIC slightly favors the more complex model, suggesting that the extra parameters may help stabilize the posterior distribution and reduce the variability between observations.

\section{Summary and conclusions}
\label{sec:conclusions}

Motivated by the persistent theoretical shortcomings and observational tensions within the standard cosmological model, in this study, we have explored the behavior of a scalar-tensor teleparallel gravity theory in which a scalar field is nonminimally coupled to torsion, allowing for a nonvanishing spatial curvature in the FRLW metric. To solve the cosmological equations, we reformulated the problem into an autonomous dynamical system and explored three distinct scenarios for the scalar field potential with vanishing, exponential and power-law forms. We thus analyzed the cosmological evolution of the given models at both background and linear perturbation levels. In our framework, the presence of a nonzero spatial curvature, in combination with the torsion-scalar coupling, introduces new dynamics that directly affect the effective DE sector. 

In particular, we investigated the DE evolution and the growth of cosmic structures by employing a numerical analysis based on a comprehensive MCMC approach involving recent cosmological observations of the Pantheon+ SN dataset combined with SH0ES Cepheid calibrations, CC measurements, and GRF data from large-scale structure probes. The parameter estimation revealed that all TDE scenarios are consistent with a nearly flat universe, although the posterior distributions for $\Omega_{k,0}$ slightly favor an open geometry. The scalar-torsion coupling parameter was robustly constrained to negative values across all models, reflecting a stable and significant deviation from minimally coupled quintessence.
The analysis of linear matter perturbations further highlights the role of the teleparallel framework in modifying the growth history of cosmic structures, demonstrating a better compatibility with observational growth rate data compared to the $\Lambda$CDM baseline.
Notably, all nonflat TDE models naturally yield higher values of $H_0$ compared to the $\Lambda$CDM model,  in closer agreement with local distance ladder measurements from SH0ES. At the same time, the $S_8$ values inferred from the TDE scenarios are systematically lower than those predicted by the Planck-$\Lambda$CDM model,
thus helping to alleviate the discrepancy with the clustering amplitude inferred from late-time large-scale structure surveys. 
This dual capability suggests that the inclusion of torsion and spatial curvature within the TDE framework provides a promising route to address two of the most persistent challenges faced by standard cosmology.

To further assess the viability of the proposed models, we performed a Bayesian model comparison using the AIC and DIC indicators. The results of our statistical analysis show that the TDE model with a vanishing scalar potential is strongly favored over $\Lambda$CDM, while the exponential and power-law potential scenarios exhibit comparable performance. Interestingly, the DIC analysis suggests that even models with extra free parameters, such as the power-law potential scenario, can be statistically competitive, indicating that the additional flexibility is meaningfully supported by the data.

In conclusion, this work provides compelling evidence that TDE models, extended to include spatial curvature, represent a robust and physically motivated alternative to the $\Lambda$CDM paradigm. In fact, the examined theoretical scenarios, while all remaining compatible with current observational constraints, not only reproduce the observed accelerated expansion but also offer an elegant pathway to mitigate key cosmological tensions.
Our results highlight the need for further exploration of teleparallel models, both from a theoretical perspective and through future high-precision cosmological surveys, as a candidate to refine our understanding of gravity and the large-scale evolution of the Universe.

A final remark concerns a possible strong coupling issue that has recently emerged in the context of cosmological perturbations in modified teleparallel gravity models  \cite{Golovnev:2018wbh,Toporensky:2021poc,Bahamonde:2022ohm}. These studies have highlighted that some additional degrees of freedom might become apparent only at higher-order perturbations or in less symmetric backgrounds.
While our work relies on the linear perturbation framework, which is commonly adopted in the literature, it is important to note that linear perturbation analyses around highly symmetric backgrounds, such as FLRW spacetimes, may not fully reveal all dynamical degrees of freedom suggested by the Hamiltonian analysis.  
Therefore,  a more comprehensive treatment of the full perturbative structure may be necessary to thoroughly understand the dynamics and stability of teleparallel gravity theories.

\acknowledgments
The authors are grateful to Daniel Blixt for the helpful discussion on this topic. R.D. acknowledges financial support from INFN -- Sezione di Roma 1, \textit{esperimento} Euclid. F.B. acknowledges INFN -- Sezione di Napoli, \textit{iniziativa specifica} GINGER. This paper is based upon work from COST Action CA21136 -- Addressing observational tensions in cosmology with systematics and fundamental physics (CosmoVerse) supported by European Cooperation in Science and Technology. 

\bibliography{references}

%merlin.mbs apsrev4-1.bst 2010-07-25 4.21a (PWD, AO, DPC) hacked
%Control: key (0)
%Control: author (8) initials jnrlst
%Control: editor formatted (1) identically to author
%Control: production of article title (-1) disabled
%Control: page (0) single
%Control: year (1) truncated
%Control: production of eprint (0) enabled
\begin{thebibliography}{95}%
\makeatletter
\providecommand \@ifxundefined [1]{%
 \@ifx{#1\undefined}
}%
\providecommand \@ifnum [1]{%
 \ifnum #1\expandafter \@firstoftwo
 \else \expandafter \@secondoftwo
 \fi
}%
\providecommand \@ifx [1]{%
 \ifx #1\expandafter \@firstoftwo
 \else \expandafter \@secondoftwo
 \fi
}%
\providecommand \natexlab [1]{#1}%
\providecommand \enquote  [1]{``#1''}%
\providecommand \bibnamefont  [1]{#1}%
\providecommand \bibfnamefont [1]{#1}%
\providecommand \citenamefont [1]{#1}%
\providecommand \href@noop [0]{\@secondoftwo}%
\providecommand \href [0]{\begingroup \@sanitize@url \@href}%
\providecommand \@href[1]{\@@startlink{#1}\@@href}%
\providecommand \@@href[1]{\endgroup#1\@@endlink}%
\providecommand \@sanitize@url [0]{\catcode `\\12\catcode `\$12\catcode
  `\&12\catcode `\#12\catcode `\^12\catcode `\_12\catcode `\%12\relax}%
\providecommand \@@startlink[1]{}%
\providecommand \@@endlink[0]{}%
\providecommand \url  [0]{\begingroup\@sanitize@url \@url }%
\providecommand \@url [1]{\endgroup\@href {#1}{\urlprefix }}%
\providecommand \urlprefix  [0]{URL }%
\providecommand \Eprint [0]{\href }%
\providecommand \doibase [0]{http://dx.doi.org/}%
\providecommand \selectlanguage [0]{\@gobble}%
\providecommand \bibinfo  [0]{\@secondoftwo}%
\providecommand \bibfield  [0]{\@secondoftwo}%
\providecommand \translation [1]{[#1]}%
\providecommand \BibitemOpen [0]{}%
\providecommand \bibitemStop [0]{}%
\providecommand \bibitemNoStop [0]{.\EOS\space}%
\providecommand \EOS [0]{\spacefactor3000\relax}%
\providecommand \BibitemShut  [1]{\csname bibitem#1\endcsname}%
\let\auto@bib@innerbib\@empty
%</preamble>
\bibitem [{\citenamefont {Berti}\ \emph {et~al.}(2015)\citenamefont {Berti}
  \emph {et~al.}}]{Berti:2015itd}%
  \BibitemOpen
  \bibfield  {author} {\bibinfo {author} {\bibfnamefont {E.}~\bibnamefont
  {Berti}} \emph {et~al.},\ }\href {\doibase 10.1088/0264-9381/32/24/243001}
  {\bibfield  {journal} {\bibinfo  {journal} {Class. Quant. Grav.}\ }\textbf
  {\bibinfo {volume} {32}},\ \bibinfo {pages} {243001} (\bibinfo {year}
  {2015})},\ \Eprint {http://arxiv.org/abs/1501.07274} {arXiv:1501.07274
  [gr-qc]} \BibitemShut {NoStop}%
\bibitem [{\citenamefont {Abbott}\ \emph {et~al.}(2016)\citenamefont {Abbott}
  \emph {et~al.}}]{LIGOScientific:2016aoc}%
  \BibitemOpen
  \bibfield  {author} {\bibinfo {author} {\bibfnamefont {B.~P.}\ \bibnamefont
  {Abbott}} \emph {et~al.} (\bibinfo {collaboration} {LIGO Scientific,
  Virgo}),\ }\href {\doibase 10.1103/PhysRevLett.116.061102} {\bibfield
  {journal} {\bibinfo  {journal} {Phys. Rev. Lett.}\ }\textbf {\bibinfo
  {volume} {116}},\ \bibinfo {pages} {061102} (\bibinfo {year} {2016})},\
  \Eprint {http://arxiv.org/abs/1602.03837} {arXiv:1602.03837 [gr-qc]}
  \BibitemShut {NoStop}%
\bibitem [{\citenamefont {Akiyama}\ \emph {et~al.}(2019)\citenamefont {Akiyama}
  \emph {et~al.}}]{EventHorizonTelescope:2019dse}%
  \BibitemOpen
  \bibfield  {author} {\bibinfo {author} {\bibfnamefont {K.}~\bibnamefont
  {Akiyama}} \emph {et~al.} (\bibinfo {collaboration} {Event Horizon
  Telescope}),\ }\href {\doibase 10.3847/2041-8213/ab0ec7} {\bibfield
  {journal} {\bibinfo  {journal} {Astrophys. J. Lett.}\ }\textbf {\bibinfo
  {volume} {875}},\ \bibinfo {pages} {L1} (\bibinfo {year} {2019})},\ \Eprint
  {http://arxiv.org/abs/1906.11238} {arXiv:1906.11238 [astro-ph.GA]}
  \BibitemShut {NoStop}%
\bibitem [{\citenamefont {Weinberg}(1989)}]{Weinberg:1988cp}%
  \BibitemOpen
  \bibfield  {author} {\bibinfo {author} {\bibfnamefont {S.}~\bibnamefont
  {Weinberg}},\ }\href {\doibase 10.1103/RevModPhys.61.1} {\bibfield  {journal}
  {\bibinfo  {journal} {Rev. Mod. Phys.}\ }\textbf {\bibinfo {volume} {61}},\
  \bibinfo {pages} {1} (\bibinfo {year} {1989})}\BibitemShut {NoStop}%
\bibitem [{\citenamefont {Padmanabhan}(2003)}]{Padmanabhan:2002ji}%
  \BibitemOpen
  \bibfield  {author} {\bibinfo {author} {\bibfnamefont {T.}~\bibnamefont
  {Padmanabhan}},\ }\href {\doibase 10.1016/S0370-1573(03)00120-0} {\bibfield
  {journal} {\bibinfo  {journal} {Phys. Rept.}\ }\textbf {\bibinfo {volume}
  {380}},\ \bibinfo {pages} {235} (\bibinfo {year} {2003})},\ \Eprint
  {http://arxiv.org/abs/hep-th/0212290} {arXiv:hep-th/0212290} \BibitemShut
  {NoStop}%
\bibitem [{\citenamefont {Joyce}\ \emph {et~al.}(2015)\citenamefont {Joyce},
  \citenamefont {Jain}, \citenamefont {Khoury},\ and\ \citenamefont
  {Trodden}}]{Joyce:2014kja}%
  \BibitemOpen
  \bibfield  {author} {\bibinfo {author} {\bibfnamefont {A.}~\bibnamefont
  {Joyce}}, \bibinfo {author} {\bibfnamefont {B.}~\bibnamefont {Jain}},
  \bibinfo {author} {\bibfnamefont {J.}~\bibnamefont {Khoury}}, \ and\ \bibinfo
  {author} {\bibfnamefont {M.}~\bibnamefont {Trodden}},\ }\href {\doibase
  10.1016/j.physrep.2014.12.002} {\bibfield  {journal} {\bibinfo  {journal}
  {Phys. Rept.}\ }\textbf {\bibinfo {volume} {568}},\ \bibinfo {pages} {1}
  (\bibinfo {year} {2015})},\ \Eprint {http://arxiv.org/abs/1407.0059}
  {arXiv:1407.0059 [astro-ph.CO]} \BibitemShut {NoStop}%
\bibitem [{\citenamefont {Ishak}(2019)}]{Ishak:2018his}%
  \BibitemOpen
  \bibfield  {author} {\bibinfo {author} {\bibfnamefont {M.}~\bibnamefont
  {Ishak}},\ }\href {\doibase 10.1007/s41114-018-0017-4} {\bibfield  {journal}
  {\bibinfo  {journal} {Living Rev. Rel.}\ }\textbf {\bibinfo {volume} {22}},\
  \bibinfo {pages} {1} (\bibinfo {year} {2019})},\ \Eprint
  {http://arxiv.org/abs/1806.10122} {arXiv:1806.10122 [astro-ph.CO]}
  \BibitemShut {NoStop}%
\bibitem [{\citenamefont {D'Agostino}\ \emph {et~al.}(2022)\citenamefont
  {D'Agostino}, \citenamefont {Luongo},\ and\ \citenamefont
  {Muccino}}]{DAgostino:2022fcx}%
  \BibitemOpen
  \bibfield  {author} {\bibinfo {author} {\bibfnamefont {R.}~\bibnamefont
  {D'Agostino}}, \bibinfo {author} {\bibfnamefont {O.}~\bibnamefont {Luongo}},
  \ and\ \bibinfo {author} {\bibfnamefont {M.}~\bibnamefont {Muccino}},\ }\href
  {\doibase 10.1088/1361-6382/ac8af2} {\bibfield  {journal} {\bibinfo
  {journal} {Class. Quant. Grav.}\ }\textbf {\bibinfo {volume} {39}},\ \bibinfo
  {pages} {195014} (\bibinfo {year} {2022})},\ \Eprint
  {http://arxiv.org/abs/2204.02190} {arXiv:2204.02190 [gr-qc]} \BibitemShut
  {NoStop}%
\bibitem [{\citenamefont {Carlip}(2001)}]{Carlip:2001wq}%
  \BibitemOpen
  \bibfield  {author} {\bibinfo {author} {\bibfnamefont {S.}~\bibnamefont
  {Carlip}},\ }\href {\doibase 10.1088/0034-4885/64/8/301} {\bibfield
  {journal} {\bibinfo  {journal} {Rept. Prog. Phys.}\ }\textbf {\bibinfo
  {volume} {64}},\ \bibinfo {pages} {885} (\bibinfo {year} {2001})},\ \Eprint
  {http://arxiv.org/abs/gr-qc/0108040} {arXiv:gr-qc/0108040} \BibitemShut
  {NoStop}%
\bibitem [{\citenamefont {Barack}\ \emph {et~al.}(2019)\citenamefont {Barack}
  \emph {et~al.}}]{Barack:2018yly}%
  \BibitemOpen
  \bibfield  {author} {\bibinfo {author} {\bibfnamefont {L.}~\bibnamefont
  {Barack}} \emph {et~al.},\ }\href {\doibase 10.1088/1361-6382/ab0587}
  {\bibfield  {journal} {\bibinfo  {journal} {Class. Quant. Grav.}\ }\textbf
  {\bibinfo {volume} {36}},\ \bibinfo {pages} {143001} (\bibinfo {year}
  {2019})},\ \Eprint {http://arxiv.org/abs/1806.05195} {arXiv:1806.05195
  [gr-qc]} \BibitemShut {NoStop}%
\bibitem [{\citenamefont {Perlmutter}\ \emph {et~al.}(1999)\citenamefont
  {Perlmutter} \emph {et~al.}}]{SupernovaCosmologyProject:1998vns}%
  \BibitemOpen
  \bibfield  {author} {\bibinfo {author} {\bibfnamefont {S.}~\bibnamefont
  {Perlmutter}} \emph {et~al.} (\bibinfo {collaboration} {Supernova Cosmology
  Project}),\ }\href {\doibase 10.1086/307221} {\bibfield  {journal} {\bibinfo
  {journal} {Astrophys. J.}\ }\textbf {\bibinfo {volume} {517}},\ \bibinfo
  {pages} {565} (\bibinfo {year} {1999})},\ \Eprint
  {http://arxiv.org/abs/astro-ph/9812133} {arXiv:astro-ph/9812133} \BibitemShut
  {NoStop}%
\bibitem [{\citenamefont {Riess}\ \emph {et~al.}(1998)\citenamefont {Riess}
  \emph {et~al.}}]{SupernovaSearchTeam:1998fmf}%
  \BibitemOpen
  \bibfield  {author} {\bibinfo {author} {\bibfnamefont {A.~G.}\ \bibnamefont
  {Riess}} \emph {et~al.} (\bibinfo {collaboration} {Supernova Search Team}),\
  }\href {\doibase 10.1086/300499} {\bibfield  {journal} {\bibinfo  {journal}
  {Astron. J.}\ }\textbf {\bibinfo {volume} {116}},\ \bibinfo {pages} {1009}
  (\bibinfo {year} {1998})},\ \Eprint {http://arxiv.org/abs/astro-ph/9805201}
  {arXiv:astro-ph/9805201} \BibitemShut {NoStop}%
\bibitem [{\citenamefont {Peebles}\ and\ \citenamefont
  {Ratra}(2003)}]{Peebles:2002gy}%
  \BibitemOpen
  \bibfield  {author} {\bibinfo {author} {\bibfnamefont {P.~J.~E.}\
  \bibnamefont {Peebles}}\ and\ \bibinfo {author} {\bibfnamefont
  {B.}~\bibnamefont {Ratra}},\ }\href {\doibase 10.1103/RevModPhys.75.559}
  {\bibfield  {journal} {\bibinfo  {journal} {Rev. Mod. Phys.}\ }\textbf
  {\bibinfo {volume} {75}},\ \bibinfo {pages} {559} (\bibinfo {year} {2003})},\
  \Eprint {http://arxiv.org/abs/astro-ph/0207347} {arXiv:astro-ph/0207347}
  \BibitemShut {NoStop}%
\bibitem [{\citenamefont {Copeland}\ \emph {et~al.}(2006)\citenamefont
  {Copeland}, \citenamefont {Sami},\ and\ \citenamefont
  {Tsujikawa}}]{Copeland:2006wr}%
  \BibitemOpen
  \bibfield  {author} {\bibinfo {author} {\bibfnamefont {E.~J.}\ \bibnamefont
  {Copeland}}, \bibinfo {author} {\bibfnamefont {M.}~\bibnamefont {Sami}}, \
  and\ \bibinfo {author} {\bibfnamefont {S.}~\bibnamefont {Tsujikawa}},\ }\href
  {\doibase 10.1142/S021827180600942X} {\bibfield  {journal} {\bibinfo
  {journal} {Int. J. Mod. Phys. D}\ }\textbf {\bibinfo {volume} {15}},\
  \bibinfo {pages} {1753} (\bibinfo {year} {2006})},\ \Eprint
  {http://arxiv.org/abs/hep-th/0603057} {arXiv:hep-th/0603057} \BibitemShut
  {NoStop}%
\bibitem [{\citenamefont {D'Agostino}(2019)}]{DAgostino:2019wko}%
  \BibitemOpen
  \bibfield  {author} {\bibinfo {author} {\bibfnamefont {R.}~\bibnamefont
  {D'Agostino}},\ }\href {\doibase 10.1103/PhysRevD.99.103524} {\bibfield
  {journal} {\bibinfo  {journal} {Phys. Rev. D}\ }\textbf {\bibinfo {volume}
  {99}},\ \bibinfo {pages} {103524} (\bibinfo {year} {2019})},\ \Eprint
  {http://arxiv.org/abs/1903.03836} {arXiv:1903.03836 [gr-qc]} \BibitemShut
  {NoStop}%
\bibitem [{\citenamefont {Riess}(2019)}]{Riess:2019qba}%
  \BibitemOpen
  \bibfield  {author} {\bibinfo {author} {\bibfnamefont {A.~G.}\ \bibnamefont
  {Riess}},\ }\href {\doibase 10.1038/s42254-019-0137-0} {\bibfield  {journal}
  {\bibinfo  {journal} {Nature Rev. Phys.}\ }\textbf {\bibinfo {volume} {2}},\
  \bibinfo {pages} {10} (\bibinfo {year} {2019})},\ \Eprint
  {http://arxiv.org/abs/2001.03624} {arXiv:2001.03624 [astro-ph.CO]}
  \BibitemShut {NoStop}%
\bibitem [{\citenamefont {Wong}\ \emph {et~al.}(2020)\citenamefont {Wong} \emph
  {et~al.}}]{H0LiCOW:2019pvv}%
  \BibitemOpen
  \bibfield  {author} {\bibinfo {author} {\bibfnamefont {K.~C.}\ \bibnamefont
  {Wong}} \emph {et~al.} (\bibinfo {collaboration} {H0LiCOW}),\ }\href
  {\doibase 10.1093/mnras/stz3094} {\bibfield  {journal} {\bibinfo  {journal}
  {Mon. Not. Roy. Astron. Soc.}\ }\textbf {\bibinfo {volume} {498}},\ \bibinfo
  {pages} {1420} (\bibinfo {year} {2020})},\ \Eprint
  {http://arxiv.org/abs/1907.04869} {arXiv:1907.04869 [astro-ph.CO]}
  \BibitemShut {NoStop}%
\bibitem [{\citenamefont {D'Agostino}\ and\ \citenamefont
  {Nunes}(2020)}]{DAgostino:2020dhv}%
  \BibitemOpen
  \bibfield  {author} {\bibinfo {author} {\bibfnamefont {R.}~\bibnamefont
  {D'Agostino}}\ and\ \bibinfo {author} {\bibfnamefont {R.~C.}\ \bibnamefont
  {Nunes}},\ }\href {\doibase 10.1103/PhysRevD.101.103505} {\bibfield
  {journal} {\bibinfo  {journal} {Phys. Rev. D}\ }\textbf {\bibinfo {volume}
  {101}},\ \bibinfo {pages} {103505} (\bibinfo {year} {2020})},\ \Eprint
  {http://arxiv.org/abs/2002.06381} {arXiv:2002.06381 [astro-ph.CO]}
  \BibitemShut {NoStop}%
\bibitem [{\citenamefont {Perivolaropoulos}\ and\ \citenamefont
  {Skara}(2022)}]{Perivolaropoulos:2021jda}%
  \BibitemOpen
  \bibfield  {author} {\bibinfo {author} {\bibfnamefont {L.}~\bibnamefont
  {Perivolaropoulos}}\ and\ \bibinfo {author} {\bibfnamefont {F.}~\bibnamefont
  {Skara}},\ }\href {\doibase 10.1016/j.newar.2022.101659} {\bibfield
  {journal} {\bibinfo  {journal} {New Astron. Rev.}\ }\textbf {\bibinfo
  {volume} {95}},\ \bibinfo {pages} {101659} (\bibinfo {year} {2022})},\
  \Eprint {http://arxiv.org/abs/2105.05208} {arXiv:2105.05208 [astro-ph.CO]}
  \BibitemShut {NoStop}%
\bibitem [{\citenamefont {Li}\ \emph {et~al.}(2023)\citenamefont {Li} \emph
  {et~al.}}]{Li:2023azi}%
  \BibitemOpen
  \bibfield  {author} {\bibinfo {author} {\bibfnamefont {S.-S.}\ \bibnamefont
  {Li}} \emph {et~al.},\ }\href {\doibase 10.1051/0004-6361/202347236}
  {\bibfield  {journal} {\bibinfo  {journal} {Astron. Astrophys.}\ }\textbf
  {\bibinfo {volume} {679}},\ \bibinfo {pages} {A133} (\bibinfo {year}
  {2023})},\ \Eprint {http://arxiv.org/abs/2306.11124} {arXiv:2306.11124
  [astro-ph.CO]} \BibitemShut {NoStop}%
\bibitem [{\citenamefont {D'Agostino}\ and\ \citenamefont
  {Nunes}(2023)}]{DAgostino:2023cgx}%
  \BibitemOpen
  \bibfield  {author} {\bibinfo {author} {\bibfnamefont {R.}~\bibnamefont
  {D'Agostino}}\ and\ \bibinfo {author} {\bibfnamefont {R.~C.}\ \bibnamefont
  {Nunes}},\ }\href {\doibase 10.1103/PhysRevD.108.023523} {\bibfield
  {journal} {\bibinfo  {journal} {Phys. Rev. D}\ }\textbf {\bibinfo {volume}
  {108}},\ \bibinfo {pages} {023523} (\bibinfo {year} {2023})},\ \Eprint
  {http://arxiv.org/abs/2307.13464} {arXiv:2307.13464 [astro-ph.CO]}
  \BibitemShut {NoStop}%
\bibitem [{\citenamefont {Di~Valentino}\ \emph {et~al.}(2025)\citenamefont
  {Di~Valentino} \emph {et~al.}}]{DiValentino:2025sru}%
  \BibitemOpen
  \bibfield  {author} {\bibinfo {author} {\bibfnamefont {E.}~\bibnamefont
  {Di~Valentino}} \emph {et~al.},\ }\href@noop {} {\  (\bibinfo {year}
  {2025})},\ \Eprint {http://arxiv.org/abs/2504.01669} {arXiv:2504.01669
  [astro-ph.CO]} \BibitemShut {NoStop}%
\bibitem [{\citenamefont {Carroll}\ \emph {et~al.}(2004)\citenamefont
  {Carroll}, \citenamefont {Duvvuri}, \citenamefont {Trodden},\ and\
  \citenamefont {Turner}}]{Carroll:2003wy}%
  \BibitemOpen
  \bibfield  {author} {\bibinfo {author} {\bibfnamefont {S.~M.}\ \bibnamefont
  {Carroll}}, \bibinfo {author} {\bibfnamefont {V.}~\bibnamefont {Duvvuri}},
  \bibinfo {author} {\bibfnamefont {M.}~\bibnamefont {Trodden}}, \ and\
  \bibinfo {author} {\bibfnamefont {M.~S.}\ \bibnamefont {Turner}},\ }\href
  {\doibase 10.1103/PhysRevD.70.043528} {\bibfield  {journal} {\bibinfo
  {journal} {Phys. Rev. D}\ }\textbf {\bibinfo {volume} {70}},\ \bibinfo
  {pages} {043528} (\bibinfo {year} {2004})},\ \Eprint
  {http://arxiv.org/abs/astro-ph/0306438} {arXiv:astro-ph/0306438} \BibitemShut
  {NoStop}%
\bibitem [{\citenamefont {Starobinsky}(2007)}]{Starobinsky:2007hu}%
  \BibitemOpen
  \bibfield  {author} {\bibinfo {author} {\bibfnamefont {A.~A.}\ \bibnamefont
  {Starobinsky}},\ }\href {\doibase 10.1134/S0021364007150027} {\bibfield
  {journal} {\bibinfo  {journal} {JETP Lett.}\ }\textbf {\bibinfo {volume}
  {86}},\ \bibinfo {pages} {157} (\bibinfo {year} {2007})},\ \Eprint
  {http://arxiv.org/abs/0706.2041} {arXiv:0706.2041 [astro-ph]} \BibitemShut
  {NoStop}%
\bibitem [{\citenamefont {Sotiriou}\ and\ \citenamefont
  {Faraoni}(2010)}]{Sotiriou:2008rp}%
  \BibitemOpen
  \bibfield  {author} {\bibinfo {author} {\bibfnamefont {T.~P.}\ \bibnamefont
  {Sotiriou}}\ and\ \bibinfo {author} {\bibfnamefont {V.}~\bibnamefont
  {Faraoni}},\ }\href {\doibase 10.1103/RevModPhys.82.451} {\bibfield
  {journal} {\bibinfo  {journal} {Rev. Mod. Phys.}\ }\textbf {\bibinfo {volume}
  {82}},\ \bibinfo {pages} {451} (\bibinfo {year} {2010})},\ \Eprint
  {http://arxiv.org/abs/0805.1726} {arXiv:0805.1726 [gr-qc]} \BibitemShut
  {NoStop}%
\bibitem [{\citenamefont {Nojiri}\ and\ \citenamefont
  {Odintsov}(2011)}]{Nojiri:2010wj}%
  \BibitemOpen
  \bibfield  {author} {\bibinfo {author} {\bibfnamefont {S.}~\bibnamefont
  {Nojiri}}\ and\ \bibinfo {author} {\bibfnamefont {S.~D.}\ \bibnamefont
  {Odintsov}},\ }\href {\doibase 10.1016/j.physrep.2011.04.001} {\bibfield
  {journal} {\bibinfo  {journal} {Phys. Rept.}\ }\textbf {\bibinfo {volume}
  {505}},\ \bibinfo {pages} {59} (\bibinfo {year} {2011})},\ \Eprint
  {http://arxiv.org/abs/1011.0544} {arXiv:1011.0544 [gr-qc]} \BibitemShut
  {NoStop}%
\bibitem [{\citenamefont {Bajardi}\ and\ \citenamefont
  {D'Agostino}(2023)}]{Bajardi:2022tzn}%
  \BibitemOpen
  \bibfield  {author} {\bibinfo {author} {\bibfnamefont {F.}~\bibnamefont
  {Bajardi}}\ and\ \bibinfo {author} {\bibfnamefont {R.}~\bibnamefont
  {D'Agostino}},\ }\href {\doibase 10.1007/s10714-023-03092-w} {\bibfield
  {journal} {\bibinfo  {journal} {Gen. Rel. Grav.}\ }\textbf {\bibinfo {volume}
  {55}},\ \bibinfo {pages} {49} (\bibinfo {year} {2023})},\ \Eprint
  {http://arxiv.org/abs/2208.02677} {arXiv:2208.02677 [gr-qc]} \BibitemShut
  {NoStop}%
\bibitem [{\citenamefont {D'Agostino}\ and\ \citenamefont
  {Luciano}(2024)}]{DAgostino:2024sgm}%
  \BibitemOpen
  \bibfield  {author} {\bibinfo {author} {\bibfnamefont {R.}~\bibnamefont
  {D'Agostino}}\ and\ \bibinfo {author} {\bibfnamefont {G.~G.}\ \bibnamefont
  {Luciano}},\ }\href {\doibase 10.1016/j.physletb.2024.138987} {\bibfield
  {journal} {\bibinfo  {journal} {Phys. Lett. B}\ }\textbf {\bibinfo {volume}
  {857}},\ \bibinfo {pages} {138987} (\bibinfo {year} {2024})},\ \Eprint
  {http://arxiv.org/abs/2408.13638} {arXiv:2408.13638 [gr-qc]} \BibitemShut
  {NoStop}%
\bibitem [{\citenamefont {Bajardi}(2024)}]{Bajardi:2024efo}%
  \BibitemOpen
  \bibfield  {author} {\bibinfo {author} {\bibfnamefont {F.}~\bibnamefont
  {Bajardi}},\ }\href {\doibase 10.1140/epjc/s10052-024-13673-x} {\bibfield
  {journal} {\bibinfo  {journal} {Eur. Phys. J. C}\ }\textbf {\bibinfo {volume}
  {84}},\ \bibinfo {pages} {1298} (\bibinfo {year} {2024})}\BibitemShut
  {NoStop}%
\bibitem [{\citenamefont {Clifton}\ \emph {et~al.}(2012)\citenamefont
  {Clifton}, \citenamefont {Ferreira}, \citenamefont {Padilla},\ and\
  \citenamefont {Skordis}}]{Clifton:2011jh}%
  \BibitemOpen
  \bibfield  {author} {\bibinfo {author} {\bibfnamefont {T.}~\bibnamefont
  {Clifton}}, \bibinfo {author} {\bibfnamefont {P.~G.}\ \bibnamefont
  {Ferreira}}, \bibinfo {author} {\bibfnamefont {A.}~\bibnamefont {Padilla}}, \
  and\ \bibinfo {author} {\bibfnamefont {C.}~\bibnamefont {Skordis}},\ }\href
  {\doibase 10.1016/j.physrep.2012.01.001} {\bibfield  {journal} {\bibinfo
  {journal} {Phys. Rept.}\ }\textbf {\bibinfo {volume} {513}},\ \bibinfo
  {pages} {1} (\bibinfo {year} {2012})},\ \Eprint
  {http://arxiv.org/abs/1106.2476} {arXiv:1106.2476 [astro-ph.CO]} \BibitemShut
  {NoStop}%
\bibitem [{\citenamefont {Koyama}(2016)}]{Koyama:2015vza}%
  \BibitemOpen
  \bibfield  {author} {\bibinfo {author} {\bibfnamefont {K.}~\bibnamefont
  {Koyama}},\ }\href {\doibase 10.1088/0034-4885/79/4/046902} {\bibfield
  {journal} {\bibinfo  {journal} {Rept. Prog. Phys.}\ }\textbf {\bibinfo
  {volume} {79}},\ \bibinfo {pages} {046902} (\bibinfo {year} {2016})},\
  \Eprint {http://arxiv.org/abs/1504.04623} {arXiv:1504.04623 [astro-ph.CO]}
  \BibitemShut {NoStop}%
\bibitem [{\citenamefont {Nojiri}\ \emph {et~al.}(2017)\citenamefont {Nojiri},
  \citenamefont {Odintsov},\ and\ \citenamefont {Oikonomou}}]{Nojiri:2017ncd}%
  \BibitemOpen
  \bibfield  {author} {\bibinfo {author} {\bibfnamefont {S.}~\bibnamefont
  {Nojiri}}, \bibinfo {author} {\bibfnamefont {S.~D.}\ \bibnamefont
  {Odintsov}}, \ and\ \bibinfo {author} {\bibfnamefont {V.~K.}\ \bibnamefont
  {Oikonomou}},\ }\href {\doibase 10.1016/j.physrep.2017.06.001} {\bibfield
  {journal} {\bibinfo  {journal} {Phys. Rept.}\ }\textbf {\bibinfo {volume}
  {692}},\ \bibinfo {pages} {1} (\bibinfo {year} {2017})},\ \Eprint
  {http://arxiv.org/abs/1705.11098} {arXiv:1705.11098 [gr-qc]} \BibitemShut
  {NoStop}%
\bibitem [{\citenamefont {Capozziello}\ \emph {et~al.}(2019)\citenamefont
  {Capozziello}, \citenamefont {D'Agostino},\ and\ \citenamefont
  {Luongo}}]{Capozziello:2019cav}%
  \BibitemOpen
  \bibfield  {author} {\bibinfo {author} {\bibfnamefont {S.}~\bibnamefont
  {Capozziello}}, \bibinfo {author} {\bibfnamefont {R.}~\bibnamefont
  {D'Agostino}}, \ and\ \bibinfo {author} {\bibfnamefont {O.}~\bibnamefont
  {Luongo}},\ }\href {\doibase 10.1142/S0218271819300167} {\bibfield  {journal}
  {\bibinfo  {journal} {Int. J. Mod. Phys. D}\ }\textbf {\bibinfo {volume}
  {28}},\ \bibinfo {pages} {1930016} (\bibinfo {year} {2019})},\ \Eprint
  {http://arxiv.org/abs/1904.01427} {arXiv:1904.01427 [gr-qc]} \BibitemShut
  {NoStop}%
\bibitem [{\citenamefont {Capozziello}\ and\ \citenamefont
  {D'Agostino}(2022)}]{Capozziello:2022wgl}%
  \BibitemOpen
  \bibfield  {author} {\bibinfo {author} {\bibfnamefont {S.}~\bibnamefont
  {Capozziello}}\ and\ \bibinfo {author} {\bibfnamefont {R.}~\bibnamefont
  {D'Agostino}},\ }\href {\doibase 10.1016/j.physletb.2022.137229} {\bibfield
  {journal} {\bibinfo  {journal} {Phys. Lett. B}\ }\textbf {\bibinfo {volume}
  {832}},\ \bibinfo {pages} {137229} (\bibinfo {year} {2022})},\ \Eprint
  {http://arxiv.org/abs/2204.01015} {arXiv:2204.01015 [gr-qc]} \BibitemShut
  {NoStop}%
\bibitem [{\citenamefont {D'Agostino}\ and\ \citenamefont
  {Nunes}(2022)}]{DAgostino:2022tdk}%
  \BibitemOpen
  \bibfield  {author} {\bibinfo {author} {\bibfnamefont {R.}~\bibnamefont
  {D'Agostino}}\ and\ \bibinfo {author} {\bibfnamefont {R.~C.}\ \bibnamefont
  {Nunes}},\ }\href {\doibase 10.1103/PhysRevD.106.124053} {\bibfield
  {journal} {\bibinfo  {journal} {Phys. Rev. D}\ }\textbf {\bibinfo {volume}
  {106}},\ \bibinfo {pages} {124053} (\bibinfo {year} {2022})},\ \Eprint
  {http://arxiv.org/abs/2210.11935} {arXiv:2210.11935 [gr-qc]} \BibitemShut
  {NoStop}%
\bibitem [{\citenamefont {Brans}\ and\ \citenamefont
  {Dicke}(1961)}]{Brans:1961sx}%
  \BibitemOpen
  \bibfield  {author} {\bibinfo {author} {\bibfnamefont {C.}~\bibnamefont
  {Brans}}\ and\ \bibinfo {author} {\bibfnamefont {R.~H.}\ \bibnamefont
  {Dicke}},\ }\href {\doibase 10.1103/PhysRev.124.925} {\bibfield  {journal}
  {\bibinfo  {journal} {Phys. Rev.}\ }\textbf {\bibinfo {volume} {124}},\
  \bibinfo {pages} {925} (\bibinfo {year} {1961})}\BibitemShut {NoStop}%
\bibitem [{\citenamefont {Horndeski}(1974)}]{Horndeski:1974wa}%
  \BibitemOpen
  \bibfield  {author} {\bibinfo {author} {\bibfnamefont {G.~W.}\ \bibnamefont
  {Horndeski}},\ }\href {\doibase 10.1007/BF01807638} {\bibfield  {journal}
  {\bibinfo  {journal} {Int. J. Theor. Phys.}\ }\textbf {\bibinfo {volume}
  {10}},\ \bibinfo {pages} {363} (\bibinfo {year} {1974})}\BibitemShut
  {NoStop}%
\bibitem [{\citenamefont {Deffayet}\ \emph {et~al.}(2011)\citenamefont
  {Deffayet}, \citenamefont {Gao}, \citenamefont {Steer},\ and\ \citenamefont
  {Zahariade}}]{Deffayet:2011gz}%
  \BibitemOpen
  \bibfield  {author} {\bibinfo {author} {\bibfnamefont {C.}~\bibnamefont
  {Deffayet}}, \bibinfo {author} {\bibfnamefont {X.}~\bibnamefont {Gao}},
  \bibinfo {author} {\bibfnamefont {D.~A.}\ \bibnamefont {Steer}}, \ and\
  \bibinfo {author} {\bibfnamefont {G.}~\bibnamefont {Zahariade}},\ }\href
  {\doibase 10.1103/PhysRevD.84.064039} {\bibfield  {journal} {\bibinfo
  {journal} {Phys. Rev. D}\ }\textbf {\bibinfo {volume} {84}},\ \bibinfo
  {pages} {064039} (\bibinfo {year} {2011})},\ \Eprint
  {http://arxiv.org/abs/1103.3260} {arXiv:1103.3260 [hep-th]} \BibitemShut
  {NoStop}%
\bibitem [{\citenamefont {Kobayashi}(2019)}]{Kobayashi:2019hrl}%
  \BibitemOpen
  \bibfield  {author} {\bibinfo {author} {\bibfnamefont {T.}~\bibnamefont
  {Kobayashi}},\ }\href {\doibase 10.1088/1361-6633/ab2429} {\bibfield
  {journal} {\bibinfo  {journal} {Rept. Prog. Phys.}\ }\textbf {\bibinfo
  {volume} {82}},\ \bibinfo {pages} {086901} (\bibinfo {year} {2019})},\
  \Eprint {http://arxiv.org/abs/1901.07183} {arXiv:1901.07183 [gr-qc]}
  \BibitemShut {NoStop}%
\bibitem [{\citenamefont {D'Agostino}\ and\ \citenamefont
  {Nunes}(2019)}]{DAgostino:2019hvh}%
  \BibitemOpen
  \bibfield  {author} {\bibinfo {author} {\bibfnamefont {R.}~\bibnamefont
  {D'Agostino}}\ and\ \bibinfo {author} {\bibfnamefont {R.~C.}\ \bibnamefont
  {Nunes}},\ }\href {\doibase 10.1103/PhysRevD.100.044041} {\bibfield
  {journal} {\bibinfo  {journal} {Phys. Rev. D}\ }\textbf {\bibinfo {volume}
  {100}},\ \bibinfo {pages} {044041} (\bibinfo {year} {2019})},\ \Eprint
  {http://arxiv.org/abs/1907.05516} {arXiv:1907.05516 [gr-qc]} \BibitemShut
  {NoStop}%
\bibitem [{\citenamefont {Hayashi}\ and\ \citenamefont
  {Shirafuji}(1979)}]{Hayashi:1979qx}%
  \BibitemOpen
  \bibfield  {author} {\bibinfo {author} {\bibfnamefont {K.}~\bibnamefont
  {Hayashi}}\ and\ \bibinfo {author} {\bibfnamefont {T.}~\bibnamefont
  {Shirafuji}},\ }\href {\doibase 10.1103/PhysRevD.19.3524} {\bibfield
  {journal} {\bibinfo  {journal} {Phys. Rev. D}\ }\textbf {\bibinfo {volume}
  {19}},\ \bibinfo {pages} {3524} (\bibinfo {year} {1979})},\ \bibinfo {note}
  {[Addendum: Phys.Rev.D 24, 3312--3314 (1982)]}\BibitemShut {NoStop}%
\bibitem [{\citenamefont {Maluf}(1994)}]{Maluf:1994ji}%
  \BibitemOpen
  \bibfield  {author} {\bibinfo {author} {\bibfnamefont {J.~W.}\ \bibnamefont
  {Maluf}},\ }\href {\doibase 10.1063/1.530774} {\bibfield  {journal} {\bibinfo
   {journal} {J. Math. Phys.}\ }\textbf {\bibinfo {volume} {35}},\ \bibinfo
  {pages} {335} (\bibinfo {year} {1994})}\BibitemShut {NoStop}%
\bibitem [{\citenamefont {Hammond}(2002)}]{Hammond:2002rm}%
  \BibitemOpen
  \bibfield  {author} {\bibinfo {author} {\bibfnamefont {R.~T.}\ \bibnamefont
  {Hammond}},\ }\href {\doibase 10.1088/0034-4885/65/5/201} {\bibfield
  {journal} {\bibinfo  {journal} {Rept. Prog. Phys.}\ }\textbf {\bibinfo
  {volume} {65}},\ \bibinfo {pages} {599} (\bibinfo {year} {2002})}\BibitemShut
  {NoStop}%
\bibitem [{\citenamefont {Krssak}\ \emph {et~al.}(2019)\citenamefont {Krssak},
  \citenamefont {van~den Hoogen}, \citenamefont {Pereira}, \citenamefont
  {B\"ohmer},\ and\ \citenamefont {Coley}}]{Krssak:2018ywd}%
  \BibitemOpen
  \bibfield  {author} {\bibinfo {author} {\bibfnamefont {M.}~\bibnamefont
  {Krssak}}, \bibinfo {author} {\bibfnamefont {R.~J.}\ \bibnamefont {van~den
  Hoogen}}, \bibinfo {author} {\bibfnamefont {J.~G.}\ \bibnamefont {Pereira}},
  \bibinfo {author} {\bibfnamefont {C.~G.}\ \bibnamefont {B\"ohmer}}, \ and\
  \bibinfo {author} {\bibfnamefont {A.~A.}\ \bibnamefont {Coley}},\ }\href
  {\doibase 10.1088/1361-6382/ab2e1f} {\bibfield  {journal} {\bibinfo
  {journal} {Class. Quant. Grav.}\ }\textbf {\bibinfo {volume} {36}},\ \bibinfo
  {pages} {183001} (\bibinfo {year} {2019})},\ \Eprint
  {http://arxiv.org/abs/1810.12932} {arXiv:1810.12932 [gr-qc]} \BibitemShut
  {NoStop}%
\bibitem [{\citenamefont {Ferraro}\ and\ \citenamefont
  {Fiorini}(2007)}]{Ferraro:2006jd}%
  \BibitemOpen
  \bibfield  {author} {\bibinfo {author} {\bibfnamefont {R.}~\bibnamefont
  {Ferraro}}\ and\ \bibinfo {author} {\bibfnamefont {F.}~\bibnamefont
  {Fiorini}},\ }\href {\doibase 10.1103/PhysRevD.75.084031} {\bibfield
  {journal} {\bibinfo  {journal} {Phys. Rev. D}\ }\textbf {\bibinfo {volume}
  {75}},\ \bibinfo {pages} {084031} (\bibinfo {year} {2007})},\ \Eprint
  {http://arxiv.org/abs/gr-qc/0610067} {arXiv:gr-qc/0610067} \BibitemShut
  {NoStop}%
\bibitem [{\citenamefont {Bengochea}\ and\ \citenamefont
  {Ferraro}(2009)}]{Bengochea:2008gz}%
  \BibitemOpen
  \bibfield  {author} {\bibinfo {author} {\bibfnamefont {G.~R.}\ \bibnamefont
  {Bengochea}}\ and\ \bibinfo {author} {\bibfnamefont {R.}~\bibnamefont
  {Ferraro}},\ }\href {\doibase 10.1103/PhysRevD.79.124019} {\bibfield
  {journal} {\bibinfo  {journal} {Phys. Rev. D}\ }\textbf {\bibinfo {volume}
  {79}},\ \bibinfo {pages} {124019} (\bibinfo {year} {2009})},\ \Eprint
  {http://arxiv.org/abs/0812.1205} {arXiv:0812.1205 [astro-ph]} \BibitemShut
  {NoStop}%
\bibitem [{\citenamefont {Linder}(2010)}]{Linder:2010py}%
  \BibitemOpen
  \bibfield  {author} {\bibinfo {author} {\bibfnamefont {E.~V.}\ \bibnamefont
  {Linder}},\ }\href {\doibase 10.1103/PhysRevD.81.127301} {\bibfield
  {journal} {\bibinfo  {journal} {Phys. Rev. D}\ }\textbf {\bibinfo {volume}
  {81}},\ \bibinfo {pages} {127301} (\bibinfo {year} {2010})},\ \bibinfo {note}
  {[Erratum: Phys.Rev.D 82, 109902 (2010)]},\ \Eprint
  {http://arxiv.org/abs/1005.3039} {arXiv:1005.3039 [astro-ph.CO]} \BibitemShut
  {NoStop}%
\bibitem [{\citenamefont {Cai}\ \emph {et~al.}(2016)\citenamefont {Cai},
  \citenamefont {Capozziello}, \citenamefont {De~Laurentis},\ and\
  \citenamefont {Saridakis}}]{Cai:2015emx}%
  \BibitemOpen
  \bibfield  {author} {\bibinfo {author} {\bibfnamefont {Y.-F.}\ \bibnamefont
  {Cai}}, \bibinfo {author} {\bibfnamefont {S.}~\bibnamefont {Capozziello}},
  \bibinfo {author} {\bibfnamefont {M.}~\bibnamefont {De~Laurentis}}, \ and\
  \bibinfo {author} {\bibfnamefont {E.~N.}\ \bibnamefont {Saridakis}},\ }\href
  {\doibase 10.1088/0034-4885/79/10/106901} {\bibfield  {journal} {\bibinfo
  {journal} {Rept. Prog. Phys.}\ }\textbf {\bibinfo {volume} {79}},\ \bibinfo
  {pages} {106901} (\bibinfo {year} {2016})},\ \Eprint
  {http://arxiv.org/abs/1511.07586} {arXiv:1511.07586 [gr-qc]} \BibitemShut
  {NoStop}%
\bibitem [{\citenamefont {Kr\v{s}\v{s}\'ak}\ and\ \citenamefont
  {Saridakis}(2016)}]{Krssak:2015oua}%
  \BibitemOpen
  \bibfield  {author} {\bibinfo {author} {\bibfnamefont {M.}~\bibnamefont
  {Kr\v{s}\v{s}\'ak}}\ and\ \bibinfo {author} {\bibfnamefont {E.~N.}\
  \bibnamefont {Saridakis}},\ }\href {\doibase 10.1088/0264-9381/33/11/115009}
  {\bibfield  {journal} {\bibinfo  {journal} {Class. Quant. Grav.}\ }\textbf
  {\bibinfo {volume} {33}},\ \bibinfo {pages} {115009} (\bibinfo {year}
  {2016})},\ \Eprint {http://arxiv.org/abs/1510.08432} {arXiv:1510.08432
  [gr-qc]} \BibitemShut {NoStop}%
\bibitem [{\citenamefont {Abedi}\ \emph {et~al.}(2018)\citenamefont {Abedi},
  \citenamefont {Capozziello}, \citenamefont {D'Agostino},\ and\ \citenamefont
  {Luongo}}]{Abedi:2018lkr}%
  \BibitemOpen
  \bibfield  {author} {\bibinfo {author} {\bibfnamefont {H.}~\bibnamefont
  {Abedi}}, \bibinfo {author} {\bibfnamefont {S.}~\bibnamefont {Capozziello}},
  \bibinfo {author} {\bibfnamefont {R.}~\bibnamefont {D'Agostino}}, \ and\
  \bibinfo {author} {\bibfnamefont {O.}~\bibnamefont {Luongo}},\ }\href
  {\doibase 10.1103/PhysRevD.97.084008} {\bibfield  {journal} {\bibinfo
  {journal} {Phys. Rev. D}\ }\textbf {\bibinfo {volume} {97}},\ \bibinfo
  {pages} {084008} (\bibinfo {year} {2018})},\ \Eprint
  {http://arxiv.org/abs/1803.07171} {arXiv:1803.07171 [gr-qc]} \BibitemShut
  {NoStop}%
\bibitem [{\citenamefont {Bahamonde}\ \emph
  {et~al.}(2023{\natexlab{a}})\citenamefont {Bahamonde}, \citenamefont
  {Dialektopoulos}, \citenamefont {Escamilla-Rivera}, \citenamefont {Farrugia},
  \citenamefont {Gakis}, \citenamefont {Hendry}, \citenamefont {Hohmann},
  \citenamefont {Levi~Said}, \citenamefont {Mifsud},\ and\ \citenamefont
  {Di~Valentino}}]{Bahamonde:2021gfp}%
  \BibitemOpen
  \bibfield  {author} {\bibinfo {author} {\bibfnamefont {S.}~\bibnamefont
  {Bahamonde}}, \bibinfo {author} {\bibfnamefont {K.~F.}\ \bibnamefont
  {Dialektopoulos}}, \bibinfo {author} {\bibfnamefont {C.}~\bibnamefont
  {Escamilla-Rivera}}, \bibinfo {author} {\bibfnamefont {G.}~\bibnamefont
  {Farrugia}}, \bibinfo {author} {\bibfnamefont {V.}~\bibnamefont {Gakis}},
  \bibinfo {author} {\bibfnamefont {M.}~\bibnamefont {Hendry}}, \bibinfo
  {author} {\bibfnamefont {M.}~\bibnamefont {Hohmann}}, \bibinfo {author}
  {\bibfnamefont {J.}~\bibnamefont {Levi~Said}}, \bibinfo {author}
  {\bibfnamefont {J.}~\bibnamefont {Mifsud}}, \ and\ \bibinfo {author}
  {\bibfnamefont {E.}~\bibnamefont {Di~Valentino}},\ }\href {\doibase
  10.1088/1361-6633/ac9cef} {\bibfield  {journal} {\bibinfo  {journal} {Rept.
  Prog. Phys.}\ }\textbf {\bibinfo {volume} {86}},\ \bibinfo {pages} {026901}
  (\bibinfo {year} {2023}{\natexlab{a}})},\ \Eprint
  {http://arxiv.org/abs/2106.13793} {arXiv:2106.13793 [gr-qc]} \BibitemShut
  {NoStop}%
\bibitem [{\citenamefont {Bajardi}\ \emph {et~al.}(2025)\citenamefont
  {Bajardi}, \citenamefont {Blixt},\ and\ \citenamefont
  {Capozziello}}]{Bajardi:2024dru}%
  \BibitemOpen
  \bibfield  {author} {\bibinfo {author} {\bibfnamefont {F.}~\bibnamefont
  {Bajardi}}, \bibinfo {author} {\bibfnamefont {D.}~\bibnamefont {Blixt}}, \
  and\ \bibinfo {author} {\bibfnamefont {S.}~\bibnamefont {Capozziello}},\
  }\href {\doibase 10.1103/PhysRevD.111.084012} {\bibfield  {journal} {\bibinfo
   {journal} {Phys. Rev. D}\ }\textbf {\bibinfo {volume} {111}},\ \bibinfo
  {pages} {084012} (\bibinfo {year} {2025})},\ \Eprint
  {http://arxiv.org/abs/2412.20592} {arXiv:2412.20592 [gr-qc]} \BibitemShut
  {NoStop}%
\bibitem [{\citenamefont {Geng}\ \emph {et~al.}(2011)\citenamefont {Geng},
  \citenamefont {Lee}, \citenamefont {Saridakis},\ and\ \citenamefont
  {Wu}}]{Geng:2011aj}%
  \BibitemOpen
  \bibfield  {author} {\bibinfo {author} {\bibfnamefont {C.-Q.}\ \bibnamefont
  {Geng}}, \bibinfo {author} {\bibfnamefont {C.-C.}\ \bibnamefont {Lee}},
  \bibinfo {author} {\bibfnamefont {E.~N.}\ \bibnamefont {Saridakis}}, \ and\
  \bibinfo {author} {\bibfnamefont {Y.-P.}\ \bibnamefont {Wu}},\ }\href
  {\doibase 10.1016/j.physletb.2011.09.082} {\bibfield  {journal} {\bibinfo
  {journal} {Phys. Lett. B}\ }\textbf {\bibinfo {volume} {704}},\ \bibinfo
  {pages} {384} (\bibinfo {year} {2011})},\ \Eprint
  {http://arxiv.org/abs/1109.1092} {arXiv:1109.1092 [hep-th]} \BibitemShut
  {NoStop}%
\bibitem [{\citenamefont {Wei}(2012)}]{Wei:2011yr}%
  \BibitemOpen
  \bibfield  {author} {\bibinfo {author} {\bibfnamefont {H.}~\bibnamefont
  {Wei}},\ }\href {\doibase 10.1016/j.physletb.2012.05.006} {\bibfield
  {journal} {\bibinfo  {journal} {Phys. Lett. B}\ }\textbf {\bibinfo {volume}
  {712}},\ \bibinfo {pages} {430} (\bibinfo {year} {2012})},\ \Eprint
  {http://arxiv.org/abs/1109.6107} {arXiv:1109.6107 [gr-qc]} \BibitemShut
  {NoStop}%
\bibitem [{\citenamefont {Geng}\ and\ \citenamefont {Wu}(2013)}]{Geng:2012vn}%
  \BibitemOpen
  \bibfield  {author} {\bibinfo {author} {\bibfnamefont {C.-Q.}\ \bibnamefont
  {Geng}}\ and\ \bibinfo {author} {\bibfnamefont {Y.-P.}\ \bibnamefont {Wu}},\
  }\href {\doibase 10.1088/1475-7516/2013/04/033} {\bibfield  {journal}
  {\bibinfo  {journal} {JCAP}\ }\textbf {\bibinfo {volume} {04}},\ \bibinfo
  {pages} {033} (\bibinfo {year} {2013})},\ \Eprint
  {http://arxiv.org/abs/1212.6214} {arXiv:1212.6214 [astro-ph.CO]} \BibitemShut
  {NoStop}%
\bibitem [{\citenamefont {D'Agostino}\ and\ \citenamefont
  {Luongo}(2018)}]{DAgostino:2018ngy}%
  \BibitemOpen
  \bibfield  {author} {\bibinfo {author} {\bibfnamefont {R.}~\bibnamefont
  {D'Agostino}}\ and\ \bibinfo {author} {\bibfnamefont {O.}~\bibnamefont
  {Luongo}},\ }\href {\doibase 10.1103/PhysRevD.98.124013} {\bibfield
  {journal} {\bibinfo  {journal} {Phys. Rev. D}\ }\textbf {\bibinfo {volume}
  {98}},\ \bibinfo {pages} {124013} (\bibinfo {year} {2018})},\ \Eprint
  {http://arxiv.org/abs/1807.10167} {arXiv:1807.10167 [gr-qc]} \BibitemShut
  {NoStop}%
\bibitem [{\citenamefont {Bajardi}\ and\ \citenamefont
  {Capozziello}(2021)}]{Bajardi:2021tul}%
  \BibitemOpen
  \bibfield  {author} {\bibinfo {author} {\bibfnamefont {F.}~\bibnamefont
  {Bajardi}}\ and\ \bibinfo {author} {\bibfnamefont {S.}~\bibnamefont
  {Capozziello}},\ }\href {\doibase 10.1142/S0219887821400028} {\bibfield
  {journal} {\bibinfo  {journal} {Int. J. Geom. Meth. Mod. Phys.}\ }\textbf
  {\bibinfo {volume} {18}},\ \bibinfo {pages} {2140002} (\bibinfo {year}
  {2021})},\ \Eprint {http://arxiv.org/abs/2101.00432} {arXiv:2101.00432
  [gr-qc]} \BibitemShut {NoStop}%
\bibitem [{\citenamefont {Yu}\ \emph {et~al.}(2018)\citenamefont {Yu},
  \citenamefont {Ratra},\ and\ \citenamefont {Wang}}]{Yu:2017iju}%
  \BibitemOpen
  \bibfield  {author} {\bibinfo {author} {\bibfnamefont {H.}~\bibnamefont
  {Yu}}, \bibinfo {author} {\bibfnamefont {B.}~\bibnamefont {Ratra}}, \ and\
  \bibinfo {author} {\bibfnamefont {F.-Y.}\ \bibnamefont {Wang}},\ }\href
  {\doibase 10.3847/1538-4357/aab0a2} {\bibfield  {journal} {\bibinfo
  {journal} {Astrophys. J.}\ }\textbf {\bibinfo {volume} {856}},\ \bibinfo
  {pages} {3} (\bibinfo {year} {2018})},\ \Eprint
  {http://arxiv.org/abs/1711.03437} {arXiv:1711.03437 [astro-ph.CO]}
  \BibitemShut {NoStop}%
\bibitem [{\citenamefont {Capozziello}\ \emph
  {et~al.}(2018{\natexlab{a}})\citenamefont {Capozziello}, \citenamefont
  {Luongo}, \citenamefont {Pincak},\ and\ \citenamefont
  {Ravanpak}}]{Capozziello:2018hly}%
  \BibitemOpen
  \bibfield  {author} {\bibinfo {author} {\bibfnamefont {S.}~\bibnamefont
  {Capozziello}}, \bibinfo {author} {\bibfnamefont {O.}~\bibnamefont {Luongo}},
  \bibinfo {author} {\bibfnamefont {R.}~\bibnamefont {Pincak}}, \ and\ \bibinfo
  {author} {\bibfnamefont {A.}~\bibnamefont {Ravanpak}},\ }\href {\doibase
  10.1007/s10714-018-2374-4} {\bibfield  {journal} {\bibinfo  {journal} {Gen.
  Rel. Grav.}\ }\textbf {\bibinfo {volume} {50}},\ \bibinfo {pages} {53}
  (\bibinfo {year} {2018}{\natexlab{a}})},\ \Eprint
  {http://arxiv.org/abs/1804.03649} {arXiv:1804.03649 [gr-qc]} \BibitemShut
  {NoStop}%
\bibitem [{\citenamefont {Bahamonde}\ \emph
  {et~al.}(2023{\natexlab{b}})\citenamefont {Bahamonde}, \citenamefont
  {Dialektopoulos}, \citenamefont {Hohmann}, \citenamefont {Levi~Said},
  \citenamefont {Pfeifer},\ and\ \citenamefont
  {Saridakis}}]{Bahamonde:2022ohm}%
  \BibitemOpen
  \bibfield  {author} {\bibinfo {author} {\bibfnamefont {S.}~\bibnamefont
  {Bahamonde}}, \bibinfo {author} {\bibfnamefont {K.~F.}\ \bibnamefont
  {Dialektopoulos}}, \bibinfo {author} {\bibfnamefont {M.}~\bibnamefont
  {Hohmann}}, \bibinfo {author} {\bibfnamefont {J.}~\bibnamefont {Levi~Said}},
  \bibinfo {author} {\bibfnamefont {C.}~\bibnamefont {Pfeifer}}, \ and\
  \bibinfo {author} {\bibfnamefont {E.~N.}\ \bibnamefont {Saridakis}},\ }\href
  {\doibase 10.1140/epjc/s10052-023-11322-3} {\bibfield  {journal} {\bibinfo
  {journal} {Eur. Phys. J. C}\ }\textbf {\bibinfo {volume} {83}},\ \bibinfo
  {pages} {193} (\bibinfo {year} {2023}{\natexlab{b}})},\ \Eprint
  {http://arxiv.org/abs/2203.00619} {arXiv:2203.00619 [gr-qc]} \BibitemShut
  {NoStop}%
\bibitem [{\citenamefont {Lewis}\ \emph {et~al.}(2000)\citenamefont {Lewis},
  \citenamefont {Challinor},\ and\ \citenamefont {Lasenby}}]{Lewis:1999bs}%
  \BibitemOpen
  \bibfield  {author} {\bibinfo {author} {\bibfnamefont {A.}~\bibnamefont
  {Lewis}}, \bibinfo {author} {\bibfnamefont {A.}~\bibnamefont {Challinor}}, \
  and\ \bibinfo {author} {\bibfnamefont {A.}~\bibnamefont {Lasenby}},\ }\href
  {\doibase 10.1086/309179} {\bibfield  {journal} {\bibinfo  {journal}
  {Astrophys. J.}\ }\textbf {\bibinfo {volume} {538}},\ \bibinfo {pages} {473}
  (\bibinfo {year} {2000})},\ \Eprint {http://arxiv.org/abs/astro-ph/9911177}
  {arXiv:astro-ph/9911177} \BibitemShut {NoStop}%
\bibitem [{\citenamefont {D'Agostino}\ \emph {et~al.}(2023)\citenamefont
  {D'Agostino}, \citenamefont {Califano}, \citenamefont {Menadeo},\ and\
  \citenamefont {Vernieri}}]{DAgostino:2023tgm}%
  \BibitemOpen
  \bibfield  {author} {\bibinfo {author} {\bibfnamefont {R.}~\bibnamefont
  {D'Agostino}}, \bibinfo {author} {\bibfnamefont {M.}~\bibnamefont
  {Califano}}, \bibinfo {author} {\bibfnamefont {N.}~\bibnamefont {Menadeo}}, \
  and\ \bibinfo {author} {\bibfnamefont {D.}~\bibnamefont {Vernieri}},\ }\href
  {\doibase 10.1103/PhysRevD.108.043538} {\bibfield  {journal} {\bibinfo
  {journal} {Phys. Rev. D}\ }\textbf {\bibinfo {volume} {108}},\ \bibinfo
  {pages} {043538} (\bibinfo {year} {2023})},\ \Eprint
  {http://arxiv.org/abs/2305.14238} {arXiv:2305.14238 [astro-ph.CO]}
  \BibitemShut {NoStop}%
\bibitem [{\citenamefont {Califano}\ \emph {et~al.}(2024)\citenamefont
  {Califano}, \citenamefont {D'Agostino},\ and\ \citenamefont
  {Vernieri}}]{Califano:2024tns}%
  \BibitemOpen
  \bibfield  {author} {\bibinfo {author} {\bibfnamefont {M.}~\bibnamefont
  {Califano}}, \bibinfo {author} {\bibfnamefont {R.}~\bibnamefont
  {D'Agostino}}, \ and\ \bibinfo {author} {\bibfnamefont {D.}~\bibnamefont
  {Vernieri}},\ }\href {\doibase 10.1103/PhysRevD.109.083520} {\bibfield
  {journal} {\bibinfo  {journal} {Phys. Rev. D}\ }\textbf {\bibinfo {volume}
  {109}},\ \bibinfo {pages} {083520} (\bibinfo {year} {2024})},\ \Eprint
  {http://arxiv.org/abs/2403.15373} {arXiv:2403.15373 [astro-ph.CO]}
  \BibitemShut {NoStop}%
\bibitem [{\citenamefont {Efstathiou}(2003)}]{Efstathiou:2003hk}%
  \BibitemOpen
  \bibfield  {author} {\bibinfo {author} {\bibfnamefont {G.}~\bibnamefont
  {Efstathiou}},\ }\href {\doibase 10.1046/j.1365-8711.2003.06940.x} {\bibfield
   {journal} {\bibinfo  {journal} {Mon. Not. Roy. Astron. Soc.}\ }\textbf
  {\bibinfo {volume} {343}},\ \bibinfo {pages} {L95} (\bibinfo {year}
  {2003})},\ \Eprint {http://arxiv.org/abs/astro-ph/0303127}
  {arXiv:astro-ph/0303127} \BibitemShut {NoStop}%
\bibitem [{\citenamefont {Lasenby}\ and\ \citenamefont
  {Doran}(2005)}]{Lasenby:2003ur}%
  \BibitemOpen
  \bibfield  {author} {\bibinfo {author} {\bibfnamefont {A.}~\bibnamefont
  {Lasenby}}\ and\ \bibinfo {author} {\bibfnamefont {C.}~\bibnamefont
  {Doran}},\ }\href {\doibase 10.1103/PhysRevD.71.063502} {\bibfield  {journal}
  {\bibinfo  {journal} {Phys. Rev. D}\ }\textbf {\bibinfo {volume} {71}},\
  \bibinfo {pages} {063502} (\bibinfo {year} {2005})},\ \Eprint
  {http://arxiv.org/abs/astro-ph/0307311} {arXiv:astro-ph/0307311} \BibitemShut
  {NoStop}%
\bibitem [{\citenamefont {Park}\ and\ \citenamefont
  {Ratra}(2019)}]{Park:2017xbl}%
  \BibitemOpen
  \bibfield  {author} {\bibinfo {author} {\bibfnamefont {C.-G.}\ \bibnamefont
  {Park}}\ and\ \bibinfo {author} {\bibfnamefont {B.}~\bibnamefont {Ratra}},\
  }\href {\doibase 10.3847/1538-4357/ab3641} {\bibfield  {journal} {\bibinfo
  {journal} {Astrophys. J.}\ }\textbf {\bibinfo {volume} {882}},\ \bibinfo
  {pages} {158} (\bibinfo {year} {2019})},\ \Eprint
  {http://arxiv.org/abs/1801.00213} {arXiv:1801.00213 [astro-ph.CO]}
  \BibitemShut {NoStop}%
\bibitem [{\citenamefont {Di~Valentino}\ \emph {et~al.}(2019)\citenamefont
  {Di~Valentino}, \citenamefont {Melchiorri},\ and\ \citenamefont
  {Silk}}]{DiValentino:2019qzk}%
  \BibitemOpen
  \bibfield  {author} {\bibinfo {author} {\bibfnamefont {E.}~\bibnamefont
  {Di~Valentino}}, \bibinfo {author} {\bibfnamefont {A.}~\bibnamefont
  {Melchiorri}}, \ and\ \bibinfo {author} {\bibfnamefont {J.}~\bibnamefont
  {Silk}},\ }\href {\doibase 10.1038/s41550-019-0906-9} {\bibfield  {journal}
  {\bibinfo  {journal} {Nature Astron.}\ }\textbf {\bibinfo {volume} {4}},\
  \bibinfo {pages} {196} (\bibinfo {year} {2019})},\ \Eprint
  {http://arxiv.org/abs/1911.02087} {arXiv:1911.02087 [astro-ph.CO]}
  \BibitemShut {NoStop}%
\bibitem [{\citenamefont {Handley}(2021)}]{Handley:2019tkm}%
  \BibitemOpen
  \bibfield  {author} {\bibinfo {author} {\bibfnamefont {W.}~\bibnamefont
  {Handley}},\ }\href {\doibase 10.1103/PhysRevD.103.L041301} {\bibfield
  {journal} {\bibinfo  {journal} {Phys. Rev. D}\ }\textbf {\bibinfo {volume}
  {103}},\ \bibinfo {pages} {L041301} (\bibinfo {year} {2021})},\ \Eprint
  {http://arxiv.org/abs/1908.09139} {arXiv:1908.09139 [astro-ph.CO]}
  \BibitemShut {NoStop}%
\bibitem [{\citenamefont {Hohmann}\ \emph {et~al.}(2018)\citenamefont
  {Hohmann}, \citenamefont {J\"arv},\ and\ \citenamefont
  {Ualikhanova}}]{Hohmann:2018rwf}%
  \BibitemOpen
  \bibfield  {author} {\bibinfo {author} {\bibfnamefont {M.}~\bibnamefont
  {Hohmann}}, \bibinfo {author} {\bibfnamefont {L.}~\bibnamefont {J\"arv}}, \
  and\ \bibinfo {author} {\bibfnamefont {U.}~\bibnamefont {Ualikhanova}},\
  }\href {\doibase 10.1103/PhysRevD.97.104011} {\bibfield  {journal} {\bibinfo
  {journal} {Phys. Rev. D}\ }\textbf {\bibinfo {volume} {97}},\ \bibinfo
  {pages} {104011} (\bibinfo {year} {2018})},\ \Eprint
  {http://arxiv.org/abs/1801.05786} {arXiv:1801.05786 [gr-qc]} \BibitemShut
  {NoStop}%
\bibitem [{\citenamefont {Golovnev}(2021)}]{Golovnev:2021lki}%
  \BibitemOpen
  \bibfield  {author} {\bibinfo {author} {\bibfnamefont {A.}~\bibnamefont
  {Golovnev}},\ }\href {\doibase 10.1088/1361-6382/ac2136} {\bibfield
  {journal} {\bibinfo  {journal} {Class. Quant. Grav.}\ }\textbf {\bibinfo
  {volume} {38}},\ \bibinfo {pages} {197001} (\bibinfo {year} {2021})},\
  \Eprint {http://arxiv.org/abs/2105.08586} {arXiv:2105.08586 [gr-qc]}
  \BibitemShut {NoStop}%
\bibitem [{\citenamefont {Blixt}\ \emph {et~al.}(2022)\citenamefont {Blixt},
  \citenamefont {Ferraro}, \citenamefont {Golovnev},\ and\ \citenamefont
  {Guzm\'an}}]{Blixt:2022rpl}%
  \BibitemOpen
  \bibfield  {author} {\bibinfo {author} {\bibfnamefont {D.}~\bibnamefont
  {Blixt}}, \bibinfo {author} {\bibfnamefont {R.}~\bibnamefont {Ferraro}},
  \bibinfo {author} {\bibfnamefont {A.}~\bibnamefont {Golovnev}}, \ and\
  \bibinfo {author} {\bibfnamefont {M.-J.}\ \bibnamefont {Guzm\'an}},\ }\href
  {\doibase 10.1103/PhysRevD.105.084029} {\bibfield  {journal} {\bibinfo
  {journal} {Phys. Rev. D}\ }\textbf {\bibinfo {volume} {105}},\ \bibinfo
  {pages} {084029} (\bibinfo {year} {2022})},\ \Eprint
  {http://arxiv.org/abs/2201.11102} {arXiv:2201.11102 [gr-qc]} \BibitemShut
  {NoStop}%
\bibitem [{\citenamefont {Golovnev}(2024)}]{Golovnev:2023qll}%
  \BibitemOpen
  \bibfield  {author} {\bibinfo {author} {\bibfnamefont {A.}~\bibnamefont
  {Golovnev}},\ }\href {\doibase 10.22323/1.455.0031} {\bibfield  {journal}
  {\bibinfo  {journal} {PoS}\ }\textbf {\bibinfo {volume} {ICPPCRubakov2023}},\
  \bibinfo {pages} {031} (\bibinfo {year} {2024})},\ \Eprint
  {http://arxiv.org/abs/2312.13858} {arXiv:2312.13858 [gr-qc]} \BibitemShut
  {NoStop}%
\bibitem [{\citenamefont {Uzan}(1999)}]{Uzan:1999ch}%
  \BibitemOpen
  \bibfield  {author} {\bibinfo {author} {\bibfnamefont {J.-P.}\ \bibnamefont
  {Uzan}},\ }\href {\doibase 10.1103/PhysRevD.59.123510} {\bibfield  {journal}
  {\bibinfo  {journal} {Phys. Rev. D}\ }\textbf {\bibinfo {volume} {59}},\
  \bibinfo {pages} {123510} (\bibinfo {year} {1999})},\ \Eprint
  {http://arxiv.org/abs/gr-qc/9903004} {arXiv:gr-qc/9903004} \BibitemShut
  {NoStop}%
\bibitem [{\citenamefont {Hohmann}(2021)}]{Hohmann:2020zre}%
  \BibitemOpen
  \bibfield  {author} {\bibinfo {author} {\bibfnamefont {M.}~\bibnamefont
  {Hohmann}},\ }\href {\doibase 10.1142/S0219887821400053} {\bibfield
  {journal} {\bibinfo  {journal} {Int. J. Geom. Meth. Mod. Phys.}\ }\textbf
  {\bibinfo {volume} {18}},\ \bibinfo {pages} {2140005} (\bibinfo {year}
  {2021})},\ \Eprint {http://arxiv.org/abs/2008.12186} {arXiv:2008.12186
  [gr-qc]} \BibitemShut {NoStop}%
\bibitem [{\citenamefont {Peebles}(1980)}]{Peebles:1980yev}%
  \BibitemOpen
  \bibfield  {author} {\bibinfo {author} {\bibfnamefont {P.~J.}\ \bibnamefont
  {Peebles}},\ }\href@noop {} {\emph {\bibinfo {title} {{The Large-Scale
  Structure of the Universe}}}}\ (\bibinfo  {publisher} {Princeton University
  Press},\ \bibinfo {year} {1980})\BibitemShut {NoStop}%
\bibitem [{\citenamefont {{Padmanabhan}}(1993)}]{1993sfu..book.....P}%
  \BibitemOpen
  \bibfield  {author} {\bibinfo {author} {\bibfnamefont {T.}~\bibnamefont
  {{Padmanabhan}}},\ }\href@noop {} {\emph {\bibinfo {title} {{Structure
  Formation in the Universe}}}}\ (\bibinfo {year} {1993})\BibitemShut {NoStop}%
\bibitem [{\citenamefont {Bahamonde}\ \emph {et~al.}(2018)\citenamefont
  {Bahamonde}, \citenamefont {B\"ohmer}, \citenamefont {Carloni}, \citenamefont
  {Copeland}, \citenamefont {Fang},\ and\ \citenamefont
  {Tamanini}}]{Bahamonde:2017ize}%
  \BibitemOpen
  \bibfield  {author} {\bibinfo {author} {\bibfnamefont {S.}~\bibnamefont
  {Bahamonde}}, \bibinfo {author} {\bibfnamefont {C.~G.}\ \bibnamefont
  {B\"ohmer}}, \bibinfo {author} {\bibfnamefont {S.}~\bibnamefont {Carloni}},
  \bibinfo {author} {\bibfnamefont {E.~J.}\ \bibnamefont {Copeland}}, \bibinfo
  {author} {\bibfnamefont {W.}~\bibnamefont {Fang}}, \ and\ \bibinfo {author}
  {\bibfnamefont {N.}~\bibnamefont {Tamanini}},\ }\href {\doibase
  10.1016/j.physrep.2018.09.001} {\bibfield  {journal} {\bibinfo  {journal}
  {Phys. Rept.}\ }\textbf {\bibinfo {volume} {775-777}},\ \bibinfo {pages} {1}
  (\bibinfo {year} {2018})},\ \Eprint {http://arxiv.org/abs/1712.03107}
  {arXiv:1712.03107 [gr-qc]} \BibitemShut {NoStop}%
\bibitem [{\citenamefont {Scolnic}\ \emph {et~al.}(2022)\citenamefont {Scolnic}
  \emph {et~al.}}]{Scolnic:2021amr}%
  \BibitemOpen
  \bibfield  {author} {\bibinfo {author} {\bibfnamefont {D.}~\bibnamefont
  {Scolnic}} \emph {et~al.},\ }\href {\doibase 10.3847/1538-4357/ac8b7a}
  {\bibfield  {journal} {\bibinfo  {journal} {Astrophys. J.}\ }\textbf
  {\bibinfo {volume} {938}},\ \bibinfo {pages} {113} (\bibinfo {year}
  {2022})},\ \Eprint {http://arxiv.org/abs/2112.03863} {arXiv:2112.03863
  [astro-ph.CO]} \BibitemShut {NoStop}%
\bibitem [{\citenamefont {Brout}\ \emph {et~al.}(2022)\citenamefont {Brout}
  \emph {et~al.}}]{Brout:2022vxf}%
  \BibitemOpen
  \bibfield  {author} {\bibinfo {author} {\bibfnamefont {D.}~\bibnamefont
  {Brout}} \emph {et~al.},\ }\href {\doibase 10.3847/1538-4357/ac8e04}
  {\bibfield  {journal} {\bibinfo  {journal} {Astrophys. J.}\ }\textbf
  {\bibinfo {volume} {938}},\ \bibinfo {pages} {110} (\bibinfo {year}
  {2022})},\ \Eprint {http://arxiv.org/abs/2202.04077} {arXiv:2202.04077
  [astro-ph.CO]} \BibitemShut {NoStop}%
\bibitem [{\citenamefont {Riess}\ \emph {et~al.}(2022)\citenamefont {Riess}
  \emph {et~al.}}]{Riess:2021jrx}%
  \BibitemOpen
  \bibfield  {author} {\bibinfo {author} {\bibfnamefont {A.~G.}\ \bibnamefont
  {Riess}} \emph {et~al.},\ }\href {\doibase 10.3847/2041-8213/ac5c5b}
  {\bibfield  {journal} {\bibinfo  {journal} {Astrophys. J. Lett.}\ }\textbf
  {\bibinfo {volume} {934}},\ \bibinfo {pages} {L7} (\bibinfo {year} {2022})},\
  \Eprint {http://arxiv.org/abs/2112.04510} {arXiv:2112.04510 [astro-ph.CO]}
  \BibitemShut {NoStop}%
\bibitem [{\citenamefont {Conley}\ \emph {et~al.}(2011)\citenamefont {Conley}
  \emph {et~al.}}]{SNLS:2011lii}%
  \BibitemOpen
  \bibfield  {author} {\bibinfo {author} {\bibfnamefont {A.}~\bibnamefont
  {Conley}} \emph {et~al.} (\bibinfo {collaboration} {SNLS}),\ }\href {\doibase
  10.1088/0067-0049/192/1/1} {\bibfield  {journal} {\bibinfo  {journal}
  {Astrophys. J. Suppl.}\ }\textbf {\bibinfo {volume} {192}},\ \bibinfo {pages}
  {1} (\bibinfo {year} {2011})},\ \Eprint {http://arxiv.org/abs/1104.1443}
  {arXiv:1104.1443 [astro-ph.CO]} \BibitemShut {NoStop}%
\bibitem [{\citenamefont {Jimenez}\ and\ \citenamefont
  {Loeb}(2002)}]{Jimenez:2001gg}%
  \BibitemOpen
  \bibfield  {author} {\bibinfo {author} {\bibfnamefont {R.}~\bibnamefont
  {Jimenez}}\ and\ \bibinfo {author} {\bibfnamefont {A.}~\bibnamefont {Loeb}},\
  }\href {\doibase 10.1086/340549} {\bibfield  {journal} {\bibinfo  {journal}
  {Astrophys. J.}\ }\textbf {\bibinfo {volume} {573}},\ \bibinfo {pages} {37}
  (\bibinfo {year} {2002})},\ \Eprint {http://arxiv.org/abs/astro-ph/0106145}
  {arXiv:astro-ph/0106145} \BibitemShut {NoStop}%
\bibitem [{\citenamefont {Capozziello}\ \emph
  {et~al.}(2018{\natexlab{b}})\citenamefont {Capozziello}, \citenamefont
  {D'Agostino},\ and\ \citenamefont {Luongo}}]{Capozziello:2017buj}%
  \BibitemOpen
  \bibfield  {author} {\bibinfo {author} {\bibfnamefont {S.}~\bibnamefont
  {Capozziello}}, \bibinfo {author} {\bibfnamefont {R.}~\bibnamefont
  {D'Agostino}}, \ and\ \bibinfo {author} {\bibfnamefont {O.}~\bibnamefont
  {Luongo}},\ }\href {\doibase 10.1016/j.dark.2018.02.002} {\bibfield
  {journal} {\bibinfo  {journal} {Phys. Dark Univ.}\ }\textbf {\bibinfo
  {volume} {20}},\ \bibinfo {pages} {1} (\bibinfo {year}
  {2018}{\natexlab{b}})},\ \Eprint {http://arxiv.org/abs/1712.04317}
  {arXiv:1712.04317 [gr-qc]} \BibitemShut {NoStop}%
\bibitem [{\citenamefont {Nesseris}\ \emph {et~al.}(2017)\citenamefont
  {Nesseris}, \citenamefont {Pantazis},\ and\ \citenamefont
  {Perivolaropoulos}}]{Nesseris:2017vor}%
  \BibitemOpen
  \bibfield  {author} {\bibinfo {author} {\bibfnamefont {S.}~\bibnamefont
  {Nesseris}}, \bibinfo {author} {\bibfnamefont {G.}~\bibnamefont {Pantazis}},
  \ and\ \bibinfo {author} {\bibfnamefont {L.}~\bibnamefont
  {Perivolaropoulos}},\ }\href {\doibase 10.1103/PhysRevD.96.023542} {\bibfield
   {journal} {\bibinfo  {journal} {Phys. Rev. D}\ }\textbf {\bibinfo {volume}
  {96}},\ \bibinfo {pages} {023542} (\bibinfo {year} {2017})},\ \Eprint
  {http://arxiv.org/abs/1703.10538} {arXiv:1703.10538 [astro-ph.CO]}
  \BibitemShut {NoStop}%
\bibitem [{\citenamefont {Wang}\ and\ \citenamefont
  {Steinhardt}(1998)}]{Wang:1998gt}%
  \BibitemOpen
  \bibfield  {author} {\bibinfo {author} {\bibfnamefont {L.-M.}\ \bibnamefont
  {Wang}}\ and\ \bibinfo {author} {\bibfnamefont {P.~J.}\ \bibnamefont
  {Steinhardt}},\ }\href {\doibase 10.1086/306436} {\bibfield  {journal}
  {\bibinfo  {journal} {Astrophys. J.}\ }\textbf {\bibinfo {volume} {508}},\
  \bibinfo {pages} {483} (\bibinfo {year} {1998})},\ \Eprint
  {http://arxiv.org/abs/astro-ph/9804015} {arXiv:astro-ph/9804015} \BibitemShut
  {NoStop}%
\bibitem [{\citenamefont {Linder}(2005)}]{Linder:2005in}%
  \BibitemOpen
  \bibfield  {author} {\bibinfo {author} {\bibfnamefont {E.~V.}\ \bibnamefont
  {Linder}},\ }\href {\doibase 10.1103/PhysRevD.72.043529} {\bibfield
  {journal} {\bibinfo  {journal} {Phys. Rev. D}\ }\textbf {\bibinfo {volume}
  {72}},\ \bibinfo {pages} {043529} (\bibinfo {year} {2005})},\ \Eprint
  {http://arxiv.org/abs/astro-ph/0507263} {arXiv:astro-ph/0507263} \BibitemShut
  {NoStop}%
\bibitem [{\citenamefont {Hastings}(1970)}]{Hastings70}%
  \BibitemOpen
  \bibfield  {author} {\bibinfo {author} {\bibfnamefont {W.~K.}\ \bibnamefont
  {Hastings}},\ }\href {http://www.jstor.org/stable/2334940} {\bibfield
  {journal} {\bibinfo  {journal} {Biometrika}\ }\textbf {\bibinfo {volume}
  {57}},\ \bibinfo {pages} {97} (\bibinfo {year} {1970})}\BibitemShut {NoStop}%
\bibitem [{\citenamefont {Aghanim}\ \emph {et~al.}(2020)\citenamefont {Aghanim}
  \emph {et~al.}}]{Planck:2018vyg}%
  \BibitemOpen
  \bibfield  {author} {\bibinfo {author} {\bibfnamefont {N.}~\bibnamefont
  {Aghanim}} \emph {et~al.} (\bibinfo {collaboration} {Planck}),\ }\href
  {\doibase 10.1051/0004-6361/201833910} {\bibfield  {journal} {\bibinfo
  {journal} {Astron. Astrophys.}\ }\textbf {\bibinfo {volume} {641}},\ \bibinfo
  {pages} {A6} (\bibinfo {year} {2020})},\ \bibinfo {note} {[Erratum:
  Astron.Astrophys. 652, C4 (2021)]},\ \Eprint
  {http://arxiv.org/abs/1807.06209} {arXiv:1807.06209 [astro-ph.CO]}
  \BibitemShut {NoStop}%
\bibitem [{\citenamefont {Akaike}(1974)}]{Akaike:1974vps}%
  \BibitemOpen
  \bibfield  {author} {\bibinfo {author} {\bibfnamefont {H.}~\bibnamefont
  {Akaike}},\ }\href {\doibase 10.1109/TAC.1974.1100705} {\bibfield  {journal}
  {\bibinfo  {journal} {IEEE Trans. Automatic Control}\ }\textbf {\bibinfo
  {volume} {19}},\ \bibinfo {pages} {716} (\bibinfo {year} {1974})}\BibitemShut
  {NoStop}%
\bibitem [{\citenamefont {Spiegelhalter}\ \emph {et~al.}(2002)\citenamefont
  {Spiegelhalter}, \citenamefont {Best}, \citenamefont {Carlin},\ and\
  \citenamefont {van~der Linde}}]{Spiegelhalter:2002yvw}%
  \BibitemOpen
  \bibfield  {author} {\bibinfo {author} {\bibfnamefont {D.~J.}\ \bibnamefont
  {Spiegelhalter}}, \bibinfo {author} {\bibfnamefont {N.~G.}\ \bibnamefont
  {Best}}, \bibinfo {author} {\bibfnamefont {B.~P.}\ \bibnamefont {Carlin}}, \
  and\ \bibinfo {author} {\bibfnamefont {A.}~\bibnamefont {van~der Linde}},\
  }\href {\doibase 10.1111/1467-9868.00353} {\bibfield  {journal} {\bibinfo
  {journal} {J. Roy. Statist. Soc. B}\ }\textbf {\bibinfo {volume} {64}},\
  \bibinfo {pages} {583} (\bibinfo {year} {2002})}\BibitemShut {NoStop}%
\bibitem [{\citenamefont {Trotta}(2008)}]{Trotta:2008qt}%
  \BibitemOpen
  \bibfield  {author} {\bibinfo {author} {\bibfnamefont {R.}~\bibnamefont
  {Trotta}},\ }\href {\doibase 10.1080/00107510802066753} {\bibfield  {journal}
  {\bibinfo  {journal} {Contemp. Phys.}\ }\textbf {\bibinfo {volume} {49}},\
  \bibinfo {pages} {71} (\bibinfo {year} {2008})},\ \Eprint
  {http://arxiv.org/abs/0803.4089} {arXiv:0803.4089 [astro-ph]} \BibitemShut
  {NoStop}%
\bibitem [{\citenamefont {Liddle}(2007)}]{Liddle:2007fy}%
  \BibitemOpen
  \bibfield  {author} {\bibinfo {author} {\bibfnamefont {A.~R.}\ \bibnamefont
  {Liddle}},\ }\href {\doibase 10.1111/j.1745-3933.2007.00306.x} {\bibfield
  {journal} {\bibinfo  {journal} {Mon. Not. Roy. Astron. Soc.}\ }\textbf
  {\bibinfo {volume} {377}},\ \bibinfo {pages} {L74} (\bibinfo {year}
  {2007})},\ \Eprint {http://arxiv.org/abs/astro-ph/0701113}
  {arXiv:astro-ph/0701113} \BibitemShut {NoStop}%
\bibitem [{\citenamefont {Kass}\ and\ \citenamefont
  {Raftery}(1995)}]{Kass:1995loi}%
  \BibitemOpen
  \bibfield  {author} {\bibinfo {author} {\bibfnamefont {R.~E.}\ \bibnamefont
  {Kass}}\ and\ \bibinfo {author} {\bibfnamefont {A.~E.}\ \bibnamefont
  {Raftery}},\ }\href {\doibase 10.1080/01621459.1995.10476572} {\bibfield
  {journal} {\bibinfo  {journal} {J. Am. Statist. Assoc.}\ }\textbf {\bibinfo
  {volume} {90}},\ \bibinfo {pages} {773} (\bibinfo {year} {1995})}\BibitemShut
  {NoStop}%
\bibitem [{\citenamefont {Golovnev}\ and\ \citenamefont
  {Koivisto}(2018)}]{Golovnev:2018wbh}%
  \BibitemOpen
  \bibfield  {author} {\bibinfo {author} {\bibfnamefont {A.}~\bibnamefont
  {Golovnev}}\ and\ \bibinfo {author} {\bibfnamefont {T.}~\bibnamefont
  {Koivisto}},\ }\href {\doibase 10.1088/1475-7516/2018/11/012} {\bibfield
  {journal} {\bibinfo  {journal} {JCAP}\ }\textbf {\bibinfo {volume} {11}},\
  \bibinfo {pages} {012} (\bibinfo {year} {2018})},\ \Eprint
  {http://arxiv.org/abs/1808.05565} {arXiv:1808.05565 [gr-qc]} \BibitemShut
  {NoStop}%
\bibitem [{\citenamefont {Toporensky}\ and\ \citenamefont
  {Tretyakov}(2022)}]{Toporensky:2021poc}%
  \BibitemOpen
  \bibfield  {author} {\bibinfo {author} {\bibfnamefont {A.~V.}\ \bibnamefont
  {Toporensky}}\ and\ \bibinfo {author} {\bibfnamefont {P.~V.}\ \bibnamefont
  {Tretyakov}},\ }\href {\doibase 10.1142/S021988782250147X} {\bibfield
  {journal} {\bibinfo  {journal} {Int. J. Geom. Meth. Mod. Phys.}\ }\textbf
  {\bibinfo {volume} {19}},\ \bibinfo {pages} {2250147} (\bibinfo {year}
  {2022})},\ \Eprint {http://arxiv.org/abs/2110.12332} {arXiv:2110.12332
  [gr-qc]} \BibitemShut {NoStop}%
\end{thebibliography}%

\begin{widetext}
\appendix*
\section{Standard FLRW cosmology}

In standard FLRW cosmology, the Hubble expansion rate is given by
\begin{equation}
    H(a)=\sqrt{\frac{\Omega_{m,0}}{a^3}+\frac{\Omega_{k,0}}{a^2}+1-\Omega_{m,0}-   \Omega_{k,0}}\ .
\end{equation}
In particular, setting $\Omega_{k,0}=0$ yields the concordance $\Lambda$CDM model. On the other hand, allowing for nonzero spatial curvature leads to the so-called $k\Lambda$CDM model. Since the standard paradigm is grounded in GR, we can set $G_\text{eff}=G$ when solving Eq.~\eqref{eq:matter_perturbations_2}.

In Table~\ref{tab:results_standard}, we report the numerical constraints up to the $2\sigma$ confidence level, obtained from the MCMC analysis of the combined background and matter perturbations measurements for the $\Lambda$CDM and $k\Lambda$CDM models. 
Also, in Fig.~\ref{fig:contours_standard}, we show the contour regions and the posterior distributions for the free parameters of the models.

\begin{table}
\begin{center}
\setlength{\tabcolsep}{1em}
\renewcommand{\arraystretch}{2.}
\begin{tabular}{c c c c c c} 
\hline
\hline
Model & $h$ & $\Omega_{m,0}$ & $\Omega_{k,0}$ & $\sigma_{8}$  \\
\hline 
$\Lambda$CDM & $0.721 \pm 0.009\,(0.017) $& $0.307\pm 0.015\,(0.029)$ & 0 & $0.766^{+0.030(0.058)}_{-0.030(0.059)}   $\\
$k\Lambda$CDM & $0.720 \pm 0.009\,(0.017)$  & $0.306 \pm 0.021\,(0.041)$ &  $0.006 \pm 0.030\,(0.064)$ & $0.763^{+0.035(0.071)}_{-0.036(0.066)}   $\\
\hline
\hline
\end{tabular}
\caption{68\% (95\%) limits on the free parameters of the standard cosmological models, obtained from the MCMC analysis on the Pantheon+SH0ES+CC+GRF data.}
 \label{tab:results_standard}
\end{center}
\end{table}

\begin{figure}
    \includegraphics[width=3.3in]{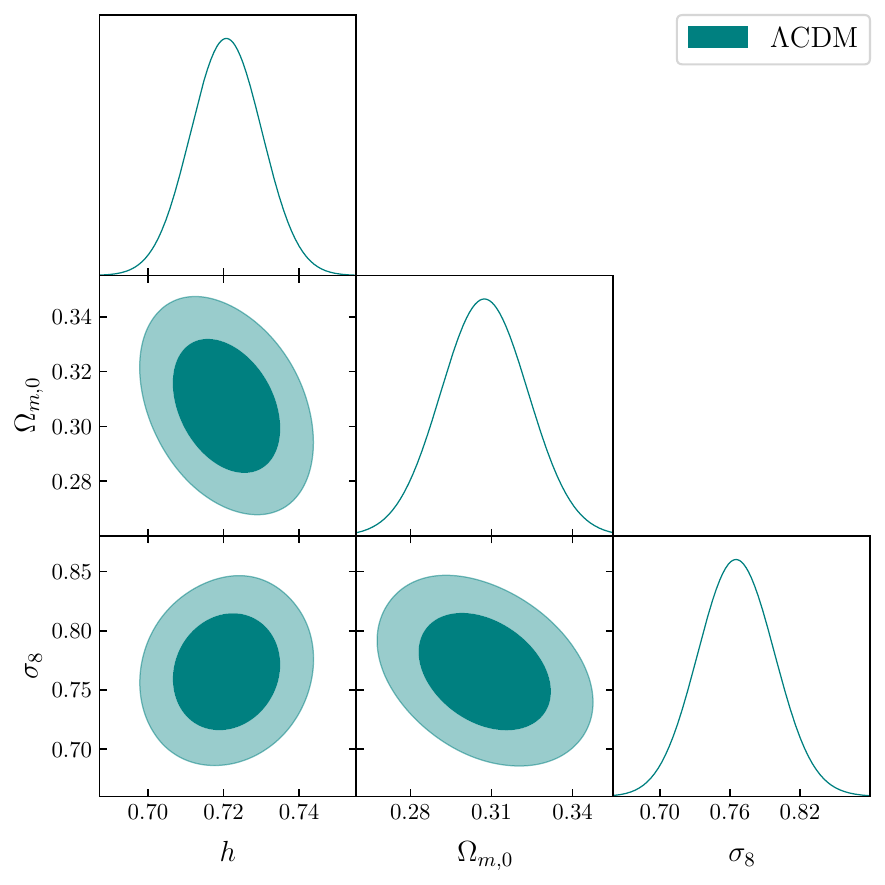}\qquad 
   \includegraphics[width=3.3in]{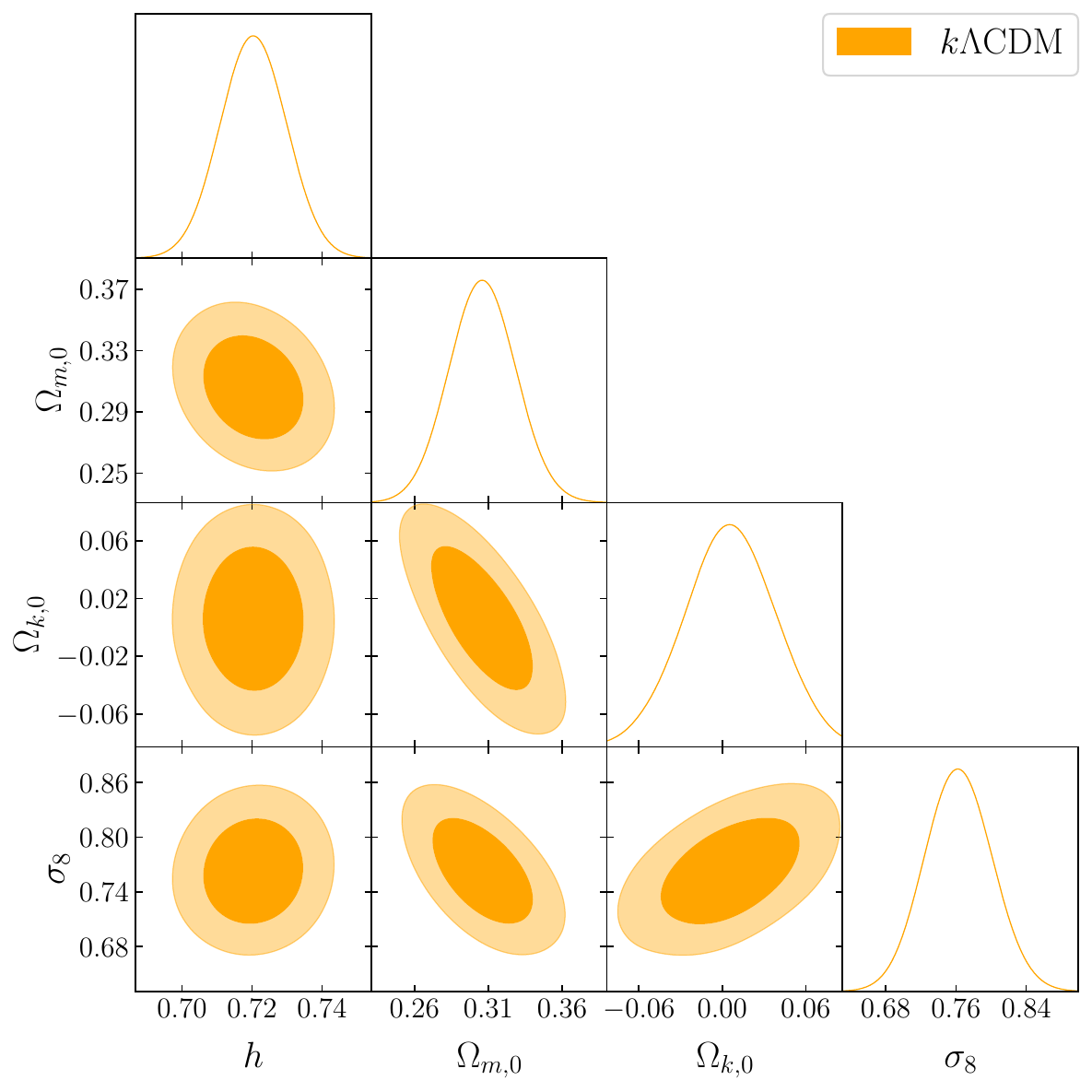}
    \caption{$68\%$ and $95\%$ contours for the standard FLRW cosmologies, obtained from the MCMC analysis on the Pantheon+SH0ES+CC+GRF data.}
    \label{fig:contours_standard}
\end{figure}

\end{widetext}

\end{document}